\documentclass[]{interact}
\usepackage{setspace}
\usepackage{xcolor}
\usepackage{epstopdf}
\usepackage{float}
\usepackage{multicol}
\usepackage{layout}

\usepackage[caption=false]{subfig}
\usepackage{amsmath}
\usepackage{soul}
\usepackage{natbib}
\bibpunct[, ]{[}{]}{,}{n}{,}{,}
\makeatletter
\def\NAT@def@citea{\def\@citea{\NAT@separator}}
\makeatother

\newcommand{\whitney}[1]{{\color{blue}#1}}

\DeclareMathOperator*{\argmin}{arg\,min} 

\theoremstyle{plain}

\theoremstyle{definition}

\theoremstyle{remark}
\usepackage{lineno}

\usepackage{booktabs}
\usepackage{longtable}
\newcommand*{\thead}[1]{%
\multicolumn{1}{c}{\bfseries\begin{tabular}{@{}c@{}}#1\end{tabular}}}
\begin{document}



\title{Estimating Value at Risk and Expected Shortfall: A Brief Review and Some New Developments}

\author{
\name{Kanon Kamronnaher\textsuperscript{a},
Andrew Bellucco\textsuperscript{b}, Whitney K.  Huang\textsuperscript{a}\thanks{CONTACT Kamronnaher. Email: kkamron@g.clemson.edu}, and Colin ~M. Gallagher\textsuperscript{a}}
\affil{\textsuperscript{a} Clemson University, School of Mathematical and Statistical Sciences, Clemson, SC;\textsuperscript{b} Credit Karma, Charlotte, NC}
}

\maketitle

\begin{abstract}

Value-at-risk (VaR) and expected shortfall (ES) are two commonly utilized metrics for quantifying financial risk. In this study, we review the widely employed Generalized Autoregressive Conditional Heteroskedasticity (GARCH) models. These models are explored with diverse distributional assumptions on innovation, including parametric, non-parametric, and `semi-parametric' that incorporates a parametric tail distribution based on extreme value theory. Additionally, we introduce  a non-parametric local linear quantile autoregression (LLQAR) with kernel weights depending on the distance between the current loss and past losses, and decreasing in the time lag.
 
To evaluate the performance of different methods for VaR and ES estimation, we employ a multi-criteria approach. This involves mean squared error assessment using simulated data, backtesting on both simulated data and US stocks, and application of the ESBootstrap test. The LLQAR method, which does not necessarily require stationarity assumptions, seems to perform better for simulated non-stationary data as well as real-world data, for estimating VaR and ES.

\end{abstract}


\section{Introduction}
\doublespacing

Value-at-Risk \citet[(VaR),][]{linsmeier2000value} is a widely used risk measure to assess the risk of different investments, which is crucial for any financial investment to ensure the safety and security of personal finances, as well as encouraging society’s trust in Financial Institutions (FIs).

VaR represents a high (low) quantile of portfolio risk, summarizing the entire risk distribution of the portfolio with a single number, thus facilitating easy comprehension.

While commonly used in quantifying financial risk, VaR has a notable limitation in
 that it does not provide any information about the size of losses beyond it. 
 Furthermore, VaR lacks coherence as a risk measure since it is non-subadditive; the portfolio's VaR might be lower than the combined VaRs of its individual components.

Expected Shortfall (ES) proposed by \cite{artzner1997thinking}
 is an alternative measure to quantify portfolio risk. It calculates
 the average of the loss distribution that  exceeds the VaR threshold, thus addressing the issue of non-subadditive inherent in VaR. 
 
In the recent Basel Committee on Banking Supervision (2016), advocated for the use of both VaR and ES to provide a more comprehensive view of market risk.



In this study, we undertake a literature review focusing on the estimation of VaR and ES using GARCH models  \citep{engle1982autoregressive, bollerslev1986generalized}  with varying assumptions regarding innovation distribution. Subsequently, we introduce a novel non-parametric approach that draws upon Quantile Autoregression models \citep[QAR,][]{koenker2006quantile}. This approach can serve as an alternative model-lean method for estimating VaR and ES.

GARCH models were introduced by \cite{engle1982autoregressive} and expanded by \cite{bollerslev1986generalized}. In these models, the variance of innovation is expressed as an autoregressive and moving average process, aiming to capture the volatility clustering feature commonly observed in financial time series. When calculating VaR and ES using GARCH models, it is essential to consider not only time-varying volatility but also the distributional assumption of the time-invariant residual term. The most common assumption is that the innovation follows a normal distribution, simplifying the estimation process and often serving as a starting point. However, empirical data frequently suggest distributions that are skewed with heavy tails, characteristics not fully captured by a normal distribution. Consequently, alternative distributions such as the Student's t-Distribution and its skewed version are also employed. An alternative approach, influenced by extreme value theory \cite{embrechts1997modelling}, involves imposing a parametric Generalized Pareto distribution for the upper tail \cite{mcneil2000estimation}. This approach can be considered a 'semi-parametric' model, providing more flexibility to capture extreme events and tail behavior.



Despite its widespread usage in estimating VaR and ES, GARCH models critically depend on model assumptions, including the parametric form of the time dynamics of volatility and the distributional assumption mentioned earlier. Therefore, there is an interest in exploring alternative, model-lean approaches.

Recently, Quantile Autoregression (QAR) introduced in \cite{koenker2006quantile} models are gaining popularity  
for modeling time series.  QAR model avoids the estimation of conditional loss distribution by directly calculating the risk estimates. Traditional least square regression can only model the conditional mean of the response variable whereas QR (QAR) allows us to model the conditional quantiles of the response variable and it does not need any distributional assumption which makes QAR more attractive to use for time series data as time series data mostly disobey the assumption of parametric form of the conditional distribution.\\
In 
this study, we combine the idea of local linear smoothing and the QAR (LLQAR) for risk estimates. 
The smoothing method of local regression was studied by  \cite{rosenblatt1956estimation}  and  \cite{parzen1962estimation} based on kernel density estimation procedure. The local linear regression smoothing technique is very popular in recent non-parametric studies. In the non-parametric study, kernel-weighted local linear fitting, together with quantile regression furnish us with information about smooth quantile curves. The idea of local linear quantile regression \cite{marasinghe2014quantile} is extended to local linear quantile autoregression. We also introduce a new weight function which depends on both past losses as well as time lags for local fitting which is not used before from the best of our knowledge. LLQAR method requires a selection of bandwidth parameter. We briefly discuss the frequently used bandwidth selection idea. Many widely used bandwidth selection procedures are established in \cite{attar2008bandwidth}. We implement a new procedure to select bandwidth based on  
cross-validation (CV).




A comprehensive comparative study is conducted to compare the estimation performance of VaR and ES using long-standing GARCH-based parametric and semi-parametric models, as well as the non-parametric LLQAR model, for risk estimation. To this end 
a simulation 
study was carried out. The methods are also applied to financial market data and analyzed their behavior. In this paper, we consider assessing the current risk of financial assets based on past and present data. We consider several time series for exposition purposes, which could be a portfolio of stocks.  Methods will be demonstrated on the S \& P 500, IBM, Apple Inc.,  and Walmart stocks. The evaluation of the performances of different methods is done by using traditional backtesting procedures for VaR and ES. 
We find that GARCH models especially the GARCH model with Extreme Value Theory (EVT) provide an acceptable forecast for both VaR and ES and unclustered VaR violations than LLQAR for stationary data in both simulation study and real-world applications. We also observe significant improvement in LLQAR for the non-stationary case that the LLQAR method is prone to yield remarkably well estimation as GARCH models. VaR and ES forecasts are also robust in both simulation studies and real-world applications for LLQAR compared to GARCH models.

The structure of the paper is as follows: Section 2 presents a literature review, while Section 3 outlines the LLQAR methods and introduces the benchmark methods applied to VaR and ES estimation. In Section 4, we evaluate the performance of the described methods through a Monte Carlo (MC) simulation study. Section 5 presents the results of various models in real-world scenarios. Lastly, Section 6 provides a summary and concludes the work.


\section{Review of methods}
  Over the years, scholars have leaned towards employing GARCH models for assessing risk in asset returns. This preference arises from the model's requirement for a minimal number of parameters for estimation, its straightforward estimation process, and its ability to capture key financial risk behaviors like volatility clustering. Moreover, GARCH models are noted for their prompt responsiveness to market movements(\cite{so2006empirical,lopez1999methods}). 
 A comparison of GARCH models for risk assessment is made on \cite{orhan2012comparison}, where they found that t-distributed GARCH is slightly better than normal, Non-Linear Power
GARCH and Non-Linear Power GARCH with a shift \cite{bollerslev1994arch}. The comparison is performed based on Kupiec \cite{kupiec1995techniques} and Christofferson \cite{christoffersen1998evaluating} tests.  
Many other researchers \cite{gao2008estimation,so2006empirical}  also conduct comparative studies on several GARCH models for evaluating financial risk. They come up with a blended conclusion that  stationary and fractionally integrated GARCH models outperform RiskMetrics and again for asymmetric fat tailed data t-GARCH performs better. 
While GARCH models are widely favored for risk assessment, they do encounter certain challenges. A significant limitation of the GARCH model is the lack of knowledge about the true distribution of innovations. Due to scatters of financial data finding an appropriate parametric distribution for innovations is not trivial. This the place where non-parametric quantile regression can exhibit it's performance as it avoids distributional assumptions and directly estimate the quantiles.
As a result, recently, many investigators are working on QAR model for using in the field of finance in different ways. 
\cite{yu1998local} developed two methods of local linear regression: one estimated quantile function by minimizing the local linear kernel version of the loss function while the other is a double kernel approach considering a kernel weighted local linear approach for estimating the conditional distribution function. \cite{el2009local}  applied local linear kernel smoothing to estimate the conditional quantile function for dependent censored data. QAR model was used to forecast the volatility of stock return in \cite{zhao2021stock} where the authors used the quantile partial autocorrelation function (QPACF) to determine the order of QAR. 
It is proved here that the out-of-sample performance of the QAR model to forecast volatility is best compared to other conventional quantile-based methods developed in \cite{taylor2005generating} and \cite{huang2012volatility} and also noticed QAR model can spell out volatility asymmetry incident. 
Details of abstraction and perquisite of QAR methods for risk analysis are clarified here  \cite{rodriguez2017five}.

Another QAR-based approach, Conditional auto regressive VaR (CAViaR) models is developed in \cite{engle2004caviar}. Their approach involves directly modeling the quantile estimates. They specified that the conditional quantiles $q_{\tau}(L_t|1:t-1)$ are auto correlated which can take the following general form:
\begin{align}
     q_{\tau}(L_t|1:t-1)=\beta_0+\sum_{i=1}^{i=p}\beta_{t-i}  q_{\tau}(L_{t-i}|1:t-i)+\sum_{j=1}^{j=q}\beta_jl|x_{t-j}|.
\end{align}
    Where, $L_t$ is the log-losses of financial time series, $\beta$'s are the unknown parameters with dimension $r=p+q+1$ need to be estimated and $l$ denotes the function of finite number of lagged variables. It has four different forms which is briefly discussed in the appendix.
In contrast, in our work, LLQAR approximates the unknown conditional quantile $q_{\tau}(L_t|1:t-1)$ by,
\begin{align}
    q_{\tau}(L_t|1:t-1) =q_{\tau}(L_{t-1}|1:t-1)+q_{\tau}'(L_{t-1}|1:t-1)(L_{t-1}-L)
\end{align}
Where, $L$ is a neighborhood point of $L_{t-1}$.
 It is observed that, Our methodology equation (2) 
is slightly different from CAViaR models equation (1) in \cite{engle2004caviar} as this models use previously estimated VaR (need to be estimated using existing methods i.e parametric approaches or historical simulation) as an explanatory variable in order to capture the dependence structure in volatility to estimate future VaR. But LLQAR does not need to use previously estimated VaR.


Since \cite{kerkhof2004backtesting} first argued that ES leads to better risk evaluation performance relative to VaR, this has been corroborated by others \cite{trindade2007approximating,liang2007risk,hurlimann2009optimisation,mittnik2009estimating}.
Recently, different forecast combining strategies are implemented by \cite{trucios2022comparison} to evaluate the performance of VaR and ES forecast for recent times cryptocurrencies. They also evaluate the performance of existing VaR and ES forecasting models such as \cite{truciosrobust}, \cite{harvey2013dynamic} and \cite{engle2004caviar} for cryptocurrency data where they found \cite{truciosrobust} and \cite{harvey2013dynamic} are the best performers.

Here \cite{nadarajah2014estimation} different formulas for estimating ES are provided only, but they
did not deliver the estimation method. Additionally, they mention certain computer software tools utilized for estimating ES.

In \cite{embrechts2005strategic}, a comparison of ES estimation methods GARCH(1,1), AR(p) and random walk model in different long term time horizon is made based on backtesting methods proposed in this work also. They found random walk model perform better than others.  A bunch of methods GARCH and APARCH with different innovations such normal,t, skewed t and General Pareto distribution (GPD) were employed and compared their performances for ES estimation based on saddle point backtest proposed by \cite{wong2008backtesting,wong2009backtesting}. They observed varied performances of different methods across different datasets. But noticeable conclusion is made here that the worst performances is given by normal innovations. An extensive comparison of ES estimation methods based on QR, CAViaR and GARCH can be read from \cite{righi2015comparison}. The corroboration that the ES leads to a better performance relative to VaR is provided by \cite{kerkhof2004backtesting}. Some research studies where ES has advantage over VaR can be read from \cite{trindade2007approximating},\cite{liang2007risk},\cite{hurlimann2009optimisation},\cite{mittnik2009estimating}. 

Our proposed methodology for estimating VaR differs from those methods described above.  We use a local linear quantile regression to allow for potential non-linear conditional quantile behavior, with dependency that decays in time.  In particular our kernel smoothing weights higher those past losses of similar magnitude to the current loss, but observations more recently in time are weighted higher as well. The method can thus be robust against non-stationary behavior.   We use our estimated VaR to find estimates of ES as described next.

\section{Statistical Methods}

In this section, we review the specifics of GARCH-based methods for estimating VaR and ES and provide details for our proposed LLQAR methods. 
 
Let ${Y_{t}}$ represent the price of an asset at time index $t$. Our focus will be on modeling the negation of the \textit{log return}, which is referred to as the \textit{log loss}. This log loss is denoted by $L_{t}$ and is defined below:

 \begin{equation}
     L_{t} = - \ln \frac{Y_{t}}{Y_{t-1}} = -\left(\ln(Y_{t} - \ln(Y_{t-1}))\right).
 \end{equation}



Let $\tau \in (0,1)$, 
then the $100(1-\tau)\%$ Value at Risk of a portfolio ($\mathrm{VaR}_\tau$) is defined as $\mathrm{VaR}_{\tau}(L_{t|1:t-1}) = \inf \{\ell \in \mathbb{R}: F_{L_{t|1:t-1}}(\ell)>1-\tau\}=F^{-1}_{L_{t|1:t-1}}(1-\tau)$, where $F_{L_{t|1:t-1}}$ denotes the conditional distribution up to the information at time $t-1$. In another word, $\mathrm{VaR}_\tau$ is the $\tau$th upper quantile of the loss distribution.

ES, also known as tail conditional expected loss, quantifies the average losses beyond the Value at Risk (VaR) and provides an indication of how much a portfolio anticipates losing. 

The expected shortfall (ES) of level $\tau$ of $L_t$ is defined by:
\begin{equation}
    \text{ES}_\tau(L_{t|1:t-1})=E(L_{t|1:t-1}|L_{t|1:t-1}>\mathrm{VaR}_\tau(L_{t|1:t-1})).
\end{equation}

 Next, we examine different statistical models and methods to estimate VaR and ES.

\subsection {Generalized Autoregressive Conditional Heteroskedastic Model (GARCH)} 
The autoregressive–moving-average model \cite{box2015time} of autoregressive order $p$ and moving average order $q$, 
ARMA($p$,$q$) 
is written as:
    \begin{align}
      L_t=\mu+\sum_{i=1}^{p}\phi_i L_{t-i}+\sum_{j=1}^{q}\theta_j\epsilon_{t-j}+\epsilon_t,
      \end{align}\\
    where, $\epsilon_t$ is the innovations of the time series, $\mu$ is the mean, $\{\phi_{i}\}_{i=1}^{p}$ is autoregressive coefficients  and $\{\theta_{j}\}_{j=1}^{q}$ is the moving average coefficients.\\

   Despite the widespread use of ARMA models for analyzing time series data, they are unable to capture heteroskedasticity behavior. This behavior refers to significant variations in variance (volatility) and clustering, commonly observed in financial time series. On the other hand, the Generalized Autoregressive Conditional Heteroskedastic (GARCH) model \cite{engle1982autoregressive,bollerslev1986generalized} is capable of accommodating a broad spectrum of dynamic conditional volatility behavior, making it particularly valuable in modeling financial time series.
   
    Specifically, GARCH model is defined as follows:
    \begin{align}
\epsilon_t=z_t\sigma_t,
 \end{align}
  \\
 where, $z_t$ is WN(0, 1).
 The innovations $\epsilon_t$ is serially uncorrelated with conditional variance $\sigma_t^2$, which vary with time taking the following form\whitney{:}
 \begin{align}
   \sigma_t^2=\omega+\sum_{i=1}^{r}\alpha_i\epsilon_{t-i}^2+\sum_{j=1}^{s}\beta_j\sigma_{t-j}^2,  
 \end{align}
 where, $\alpha_i$ and $\beta_j$ are the moving average and autoregressive coefficients analogous to those in ARMA but for the second order structure, respectively.
 

As the conditional distribution of $L_{t}$,
 various assumptions can be applied to $z_t$, encompassing parametric approaches like Gaussian or Student's t distributions, non-parametric methods, or semi-parametric approaches involving the upper tail modeled with the Generalized Pareto Distribution (GPD).

 If the distribution of innovations ($z_t$) is assumed to be normal, the corresponding model is referred to as nGARCH (normal GARCH). The estimation of Value at Risk (VaR) and Expected Shortfall (ES) for nGARCH is performed as follows:
\begin{align}
  VaR_\tau (t)=\sigma_t \Phi^{-1
}{(\tau)},  
\end{align}
 where, $\Phi^{-1}{(\tau)}$ is the standard normal (upper) quantile function \cite{manganelli2001value} and
 \begin{align}
   ES_\tau (t)=\sigma_t \frac{\phi(\Phi^{-1
}{(\tau)})}{\tau},  
 \end{align}
where, $\phi$ is the standard normal density function \cite{nadarajah2014estimation}.

If the distribution of innovations ($z_t$) is assumed to follow a Student's t distribution, the corresponding model is referred to as tGARCH (Student's t GARCH). The estimation of Value at Risk (VaR) and Expected Shortfall (ES) for tGARCH is carried out as follows:
\begin{align}
  VaR_\tau (t)=\sigma_t (\frac{\nu}{\nu-2} )T^{-1
}{(\tau)},  
\end{align}
 where, $T^{-1}{(\tau)}$ is the standard (upper) t quantile function and $\nu$ is the degrees of freedom of t distribution \cite{manganelli2001value} and
 \begin{align}
   ES_\tau (t)=\sigma_t \frac{\nu-(T^{-1}(\tau))^2}{\nu-1}\frac{t(T^{-1}(\tau))}{\tau},  
 \end{align}
where, $t$ is the standard t density function \cite{nadarajah2014estimation} \cite{broda2011expected}.

In the DFGARCH(1,1) model, the assumption is made that it follows the same GARCH(1,1) structure, but without imposing any specific distributional assumption on the innovations. In this scenario, the expressions for Value at Risk (VaR) and Expected Shortfall (ES) are as follows:
\begin{align}
    VaR_\tau (t)=\sigma_t q_{emp}
{(\tau)},
\end{align} where, $q_{emp}{(\tau)}$ is the   empirical upper
quantile of the standardized residuals, and
\begin{align}
    ES_\tau (t)=\frac{\sigma_t}{\lceil{(1-\tau) n}\rceil} \sum_{k=1}^{k=\lceil{(1-\tau) n}\rceil}e_{(k)},
\end{align}
where, $e_{(1)}\leq e_{(2)}\leq .....\leq e_{(n)}$ be the order statistics, in increasing order of the standardized residuals from the GARCH(1,1) model and ${\lceil{. }\rceil}$ is the ceiling function.

If the upper tail of the innovations is assumed to follow the Generalized Pareto Distribution (GPD), the resulting model is often referred to as GPD-GARCH. \cite{mcneil2000estimation}.

In the framework of Extreme Value Analysis (EVA), as described by Pickands (1975); Balkema \& de Haan (1974) and \cite{pickands1975statistical} the theorem states that under certain conditions, Peak-Over-Threshold (POT) models for log-loss ($L_t$) exceedances, considering a sufficiently high threshold $u$, can be modeled by the Generalized Pareto Distribution (GPD). The expression for this modeling is as follows:
\begin{align*}
    G_{(\zeta,\psi)}(l_t)= \begin{cases} 
      1-(1+\frac{\zeta l_t}{\psi})^\frac{-1}{\zeta} , & \zeta \neq 0\\
      1-\exp(\frac{-l_t}{\psi}) , & \zeta = 0\\
      \end{cases}
\end{align*}
 where, $\zeta \in (-\infty,\infty)$ is the shape parameter, and $\psi\in (0,\infty)$ is the scale parameter.\\
 Indeed, there are various methods to estimate the parameters of the Generalized Pareto Distribution (GPD), and one common approach is the Maximum Likelihood Estimate (MLE).  For threshold $u$, coverage  level $p=1-\tau$, the quantile  is estimated by 
\begin{align*}
    \hat{q}_{GPD}{(\tau)}=u+\frac{-\psi}{\zeta}(1-(\frac{n}{N_u}(1-p))^\zeta),
\end{align*}
where, $n$ is the total number of observations, and $N_u$ is the number of excesses over threshold.\\
In this framework, VaR can be estimated by \cite{marimoutou2009extreme}, 
\begin{center}
$VaR_\tau{(t)}=\sigma_t\hat{q}_{GPD}{(\tau)}$,
 \end{center}
also we obtain the corresponding ES estimate as \cite{mcneil1999extreme} , 
\begin{center}
$ES_\tau(t)=\sigma_t(\frac{\hat{q}_{GPD}{(\tau)}}{1-\zeta}+\frac{\psi-\zeta u}{1-\zeta})$,
 \end{center}
where, $\sigma_t$ is the conditional SD at time t.\\
\\
As a result, we come by the following GARCH methods for assessing VaR and ES ,

\begin{center}
    \begin{tabular}{ l | l | l  }
      \hline
      \hline
      \thead{Model} &
      \thead{$VaR_\tau(t)$} & \thead{$ES_\tau(t)$}   \\
      \hline
      
       nGARCH  & $\sigma_t \Phi^{-1
}{(\tau)}$ & $\sigma_t \frac{\phi(\Phi^{-1
}{(\tau)})}{\tau}$ \\
 &\\
        tGARCH& $ \sigma_t (\frac{\nu}{\nu-2} )t^{-1
}{(\tau)}$ & $\sigma_t \frac{\nu-(T^{-1}(\tau))^2}{\nu-1}\frac{t(T^{-1}(\tau))}{\tau}$ \\
& \\
    DFGARCH & $\sigma_t \hat{q}_{emp}
{(\tau)}$ & $\frac{\sigma_t}{\lceil{(1-\tau) n\rceil}} \sum_{k=1}^{k=\lceil{(1-\tau) n}\rceil}e_{(k)}$\\
      &\\
      GPD-GARCH & $\sigma_t\hat{q}_{GPD}(\tau)$ & $\sigma_t\frac{(\hat{q}_{GPD}(\tau)}{1-\zeta}+\frac{\psi-\zeta u}{1-\zeta})$.\\ &
      \\
      \hline
    \end{tabular}
  \end{center}

\subsection{Quantile Autoregression (QAR) method} \label{QAR}
 In the context of financial modeling, GARCH  models are commonly used to capture time-varying volatility in financial time series data. However, GARCH-based parametric models can lead to misleading or incorrect inferences because of a mischaracterized loss distribution \cite{wang2002asymptotic}. To address these issues, researchers often consider alternative models that relax distributional assumptions or use non-parametric approaches. 
The QAR model, introduced by Koenker and Machado in their work from \cite{koenker2006quantile} is highlighted for its distinctive characteristic of not requiring any specific distributional assumption. This quality contributes to the robustness of the model, making it less susceptible to bias arising from distributional misspecifications. The key advantage of the QAR model lies in its ability to describe the conditional distribution of variables at various quantile levels. This flexibility enables the model to capture and accommodate heavy-tailed behavior in log-losses.

As the QAR model directly computes the quantile of the log losses, the $\tau^{th}$ conditional upper quantile of $L_t$ is defined as follows:

\begin{align}
    q_{\tau}(L_{t|1:t-1}|F_{t-1}) =\theta_0 {(\tau)}+\theta_1 {(\tau)}L_{t-1}+\theta_2 {(\tau)}L_{t-2}+....+\theta_p {(\tau)}L_{t-p},
\end{align}
where, $\tau \in (0,1)$ is chosen to be small (i.e $\tau=0.05$ or $\tau=0.01$), and $F_{t-1}$  is the information set available at time $t-1$. According to the definition of VaR, $VaR_\tau{(t)}$ is defined as follows: 
\begin{align}
    Pr(L_{t|1:t-1}>VaR_\tau {(t)}|F_{t-1})=\tau,
\end{align}
where, $L_{t|1:t-1}$ denote log-losses of some portfolio. 
Consider the case of $p=1$ which is called QAR(1) model and is defined as follows:
\begin{align}
    q_{\tau}(L_{t|1:t-1}|F_{t-1}) =\theta_0 {(\tau)}+\theta_1 {(\tau)} L_{t-1}.
\end{align}
The estimation procedure for the QAR (Quantile Autoregressive) model is akin to the estimation process used in standard linear quantile regression. Linear quantile regression, in general, extends traditional linear regression by modeling the conditional quantiles of the response variable. Similarly, the QAR model extends this concept to time series data, incorporating autoregressive components. We can estimate both $\theta_0 {(\tau)}$ and  $\theta_1 {(\tau)}$ by solving the following minimization problem:
\begin{align}
    \min\limits_{\theta_0,\theta_1}\sum_{t=1}^{t=T}\rho_\tau(L_{t|1:t-1}-\theta_0 {(\tau)}-\theta_1 {(\tau)} L_{t-1}),
\end{align}
where, $T$ is the total number of observations, $\rho_\tau(x)=x(\tau-1_{[x\leq 0]})$, and the indicator function, 
\begin{align*}
    1_{\left[x\leq 0\right]}= \begin{cases} 
      1 , & x\leq 0,\\
      0 , & \text{otherwise}.\\
      \end{cases}
\end{align*}
\subsubsection{Local Linear Quantile Autoregression(LLQAR)}
The Local Linear Quantile Autoregression (LLQAR) is a significant non-parametric tool that extends the concept of local regression smoothing. It combines elements from the work of  \cite{rosenblatt1956estimation} and \cite{parzen1962estimation} on local regression smoothing with the ideas introduced in quantile autoregression by  \cite{koenker2006quantile}. It leverages local regression smoothing techniques to estimate the conditional quantiles adaptively around each data point, considering the autoregressive structure of the time series.
The basic concept of local linear fitting involves approximating the unknown quantile autoregression function $m_\tau(L_{t})$ in the neighborhood of $L$, $L$ is close to $L_t$,
\begin{align}
    m_\tau(L)=m_\tau(L_{t})+{m_\tau(L_{t})}'(L_t-L)=:\hat{Y}^T\beta_\tau,
\end{align}
where, $m_{\tau}^{'}$ is the first derivative of $m_{\tau}$, $\hat{Y}=(1,L_t-L)^T$, and $\beta_\tau=(\alpha_\tau=m_\tau,\beta_{1\tau}={m_\tau}')$. The literature from Section \ref{QAR} advises that the estimate $\beta_\tau$ is solved by using the following objective function:
\begin{align}
    \argmin\limits_{\beta_\tau}\sum_{t=1}^{t=T}\rho_\tau{\left(L-\hat{Y}^T\beta_\tau\right)}w_t(L_{t}),
\end{align}
where, $\rho_\tau$ is the traditional check function defined in Section \ref{QAR}, and $w_t(L_{t})=K\left(\frac{(L_t-L)(T-(t:(T-1)))/(T-t)}{h}\right)$ with $K$ is a Kernel function defined on $R^n$ and a bandwidth parameter $h$. Here the kernel function depends on the distance between the current loss L and the past Losses $L_t$ as well as
their corresponding time lag value. The averaging over the time lag will sufficiently smooth the data. The bandwidth selection methodologies are discussed next.
\subsubsection{Bandwidth Selection}

Bandwidth selection plays a crucial role in smoothing as well as reducing bias-variance trade-offs. Although, there is plenty of literature has been carried out over time about selecting bandwidth in the mean regression \cite{cai2000functional}. But there is no stable idea at hand in bandwidth selection for the QR approach.  Thus, the existing method of bandwidth selection has been used in the QR discipline. Here, some very often used bandwidth formula is discussed below briefly.

One of the most utilized data-driven methods is Cross-Validation (CV). CV uncovers prediction error by using the following leave-one-out CV statistics:
    \begin{align}
        CV(h)=\argmin\limits_h\sum\limits_{t=1}^{t=T}\rho_\tau\left(L_t-\hat{q}^{(t)}_{L_t}(\tau|F_{t-1})\right).
    \end{align}
     
   But it has one issue, due to a relatively low convergence rate, i.e., $o(n)^{-\frac{1}{10}}$, it needs high computational cost. More details about CV can be found in \cite{abberger1998cross}.

    Some other commonly used bandwidth selection formulas for smoothing quantile regression parameters are bellowed:

 1.  $h_\tau=h_{mean}\left(\frac{\tau(1-\tau)}{\big(\phi(\Phi^{-1}(\tau))\big)^{2}}\right)^{\frac{1}{5}}, $
    
    Yu and Jones \cite{yu1998local}.
   The classical technique of Ruppert et al. \cite{ruppert1995effective} can easily be used to find the $h_{mean}$.

     2.   $h_{HS}=n^{-\frac{1}{2}}z_\tau^{\frac{2}{3}}\left[\frac{1.5\phi^2(\Phi^{-1}(\tau))}{2{(\Phi^{-1}(\tau))}^2+1}\right],$
    
    where $z_\tau $ satisfies  $\Phi(z_\tau)=1-\frac{\tau}{2}$ Hall and Sheather  \cite{hall1988distribution}.

     3.  $ h_{B}=n^{-\frac{1}{5}}\left[\frac{4.5\phi^4(\Phi^{-1}(\tau))}{{(2{(\Phi^{-1}(\tau))}^2+1)}^2}\right]$ Bonfinger  \cite{bofingeb1975estimation}.

In this present work for the LLQAR approach, we checked all the bandwidth.   
Although, CV is very expensive and very slowly converged compared to computational cost, but it outperforms LLQAR. In our work, we establish a bandwidth selection policy called Quantile Cross-Validation (QCV) bandwidth. Here we use CV methodology in the first window of every time series to choose the optimal quantile ($q_{opt}$), then use it to calculate the bandwidth as,
\begin{center}
    $h=q_{opt}$ th quantile of $(|(L_t-L)(T-(t:(T-1)))/(T-t)|)$.
\end{center}
We also recalculate bandwidth using this bandwidth idea in each window for every time series but it does not improve the VaR and ES estimate than calculating one using the first window. Also, it has a very low convergence rate and makes it computationally very expensive.\\
Also, we found another bandwidth that has a similar performance as QCV but converges rate is higher than QCV which usher us to use this bandwidth. This bandwidth is defined as follows:\\
\begin{center}
    $h=(4/(3*n))^{(1/5)}*IQR (|(L_t-L)(T-(t:(T-1)))/(T-t)|)$,
\end{center}
where, $n$ is the sample size.\\
In this section, we have provided a concise overview of the Value at Risk (VaR) and Expected Shortfall (ES) estimation concepts, highlighting both parametric and non-parametric statistical methods.

\subsubsection{VaR and ES evaluation methods}
To assess the performance of Value at Risk (VaR) and Expected Shortfall (ES) models, traditional backtesting methods are employed. This statistical procedure involves comparing the estimated VaR and ES values with the actual losses. In this study, the backtesting technique proposed by \cite{kupiec1995techniques} and \cite {christoffersen1998evaluating} is utilized, which assesses both the number of exceedances through the Unconditional Coverage test (UC) and their independence through the Conditional Coverage test (CC).

Compared to VaR, evaluating ES is more challenging as ES is not elicitable. To assess ES estimates, this study adopts methods introduced and discussed in more detail by \cite{mcneil2000estimation} specifically using the ES Bootstrap test. This technique provides a rigorous evaluation of ES estimates by considering the entire distribution of losses. We also introduce and apply an assessment method for ES based on the work of \cite{embrechts2005strategic}. The combined use of these backtesting methods provides a comprehensive evaluation framework for both VaR and ES models, ensuring a thorough examination of their performance and reliability in capturing extreme losses.

\section{Simulation Study}

\subsection{Setup}
 In this phase, a simulation study is undertaken to assess the performance of Value at Risk (VaR) and Expected Shortfall (ES) estimation. The evaluation encompasses GARCH-based models that have been considered, along with the version of Local Linear Quantile Autoregression (LLQAR) implemented in this study.
 
 The data generating mechanism is chosen based on the exponential GARCH model \citep[(eGARCH),][]{nelson1991conditional} with student t innovation distribution, which removes the symmetric innovation distributional assumption of GARCH, and therefore provides a reasonable ``ground truth'' for the comparison. The parameters of eGARCH are chosen by fitting it to Apple Stock from December 2 2014 to November 9 2022. The figure below represents an example of simulated time series with True VaR and True ES.   

 \begin{figure}[H]
    \centering
    \includegraphics[width=3.4in]{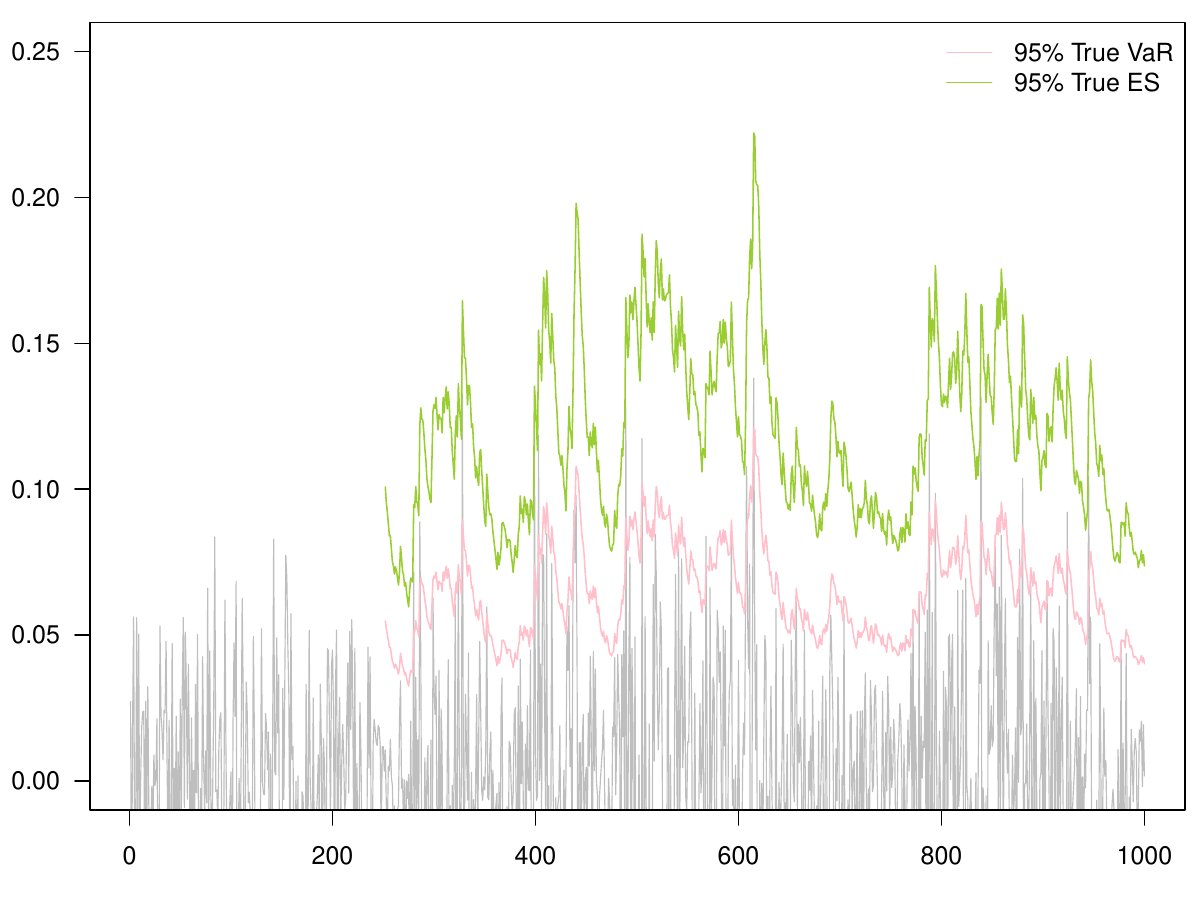}
     \caption{ One step ahead 95\%  True VaR and 95\%  True ES.}\label{fig1}
\end{figure}

In our Monte Carlo experiment, 1000 observations are simulated (roughly equivalent to four years of data) for each realization of the chosen eGARCH model and 100 realizations are generated in order to examine the sampling properties of various performance metrics regarding VaR and ES among the models we considered here (GARCH(1,1) with different innovation distribution assumptions (i.e., \texttt{nGARCH}, \texttt{tGARCH}, \texttt{DFGARCH}, and \texttt{gpdGARCH}) and \texttt{LLQAR}).

In addition to the eGARCH model used to generate simulated data, we also create two more scenarios by modifying the innovation $\sigma_{t}=a_{t}\epsilon_{t}$ with a non-negative time-varying multiplicative factor $\gamma_{t}$ and hence $\sigma_{t}=\gamma_{t}a_{t}\epsilon_{t}$, therefore the aforementioned scenario corresponds to the case where $\gamma_{t}=1$. Specifically, we consider the following choices of $\gamma_t$: (1) $\gamma_t$ is a step function, and (2) $\gamma_t$ is a smooth function, both of which fluctuate around 1 to introduce different kinds of non-stationarity of the eGARCH model. 

\begin{figure}[H]
    \centering
    \includegraphics[width=3.4in]{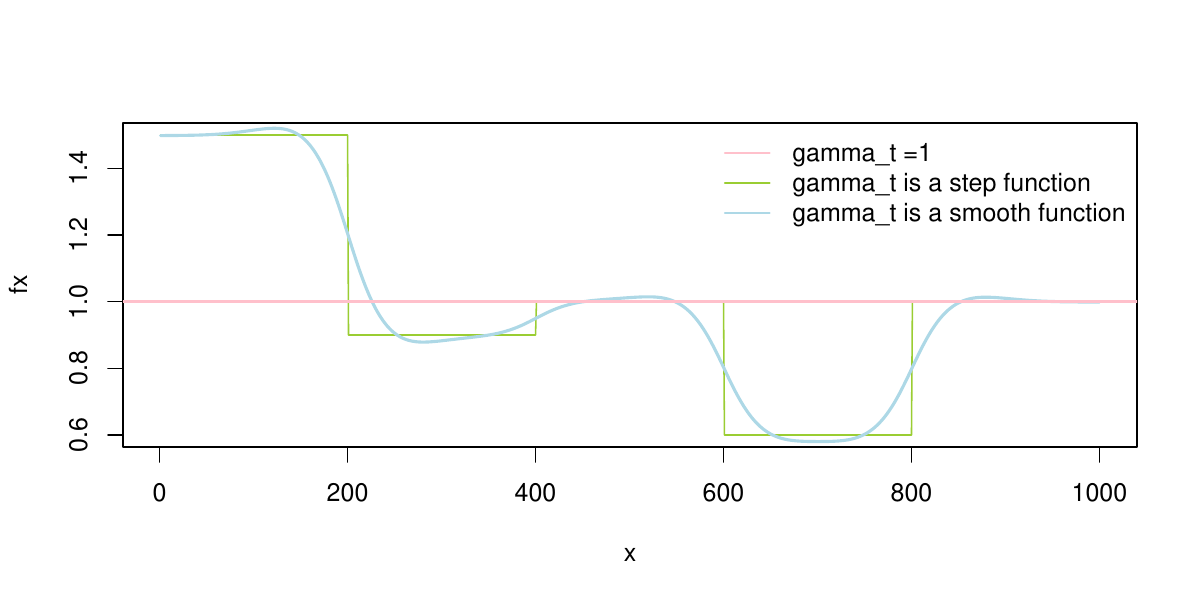}
     \caption{ Plot of $\gamma_t$ 's  used here to generate simulations.}\label{fig2}
\end{figure}

\begin{figure}[H]
    \centering
    \includegraphics[width=3.4in]{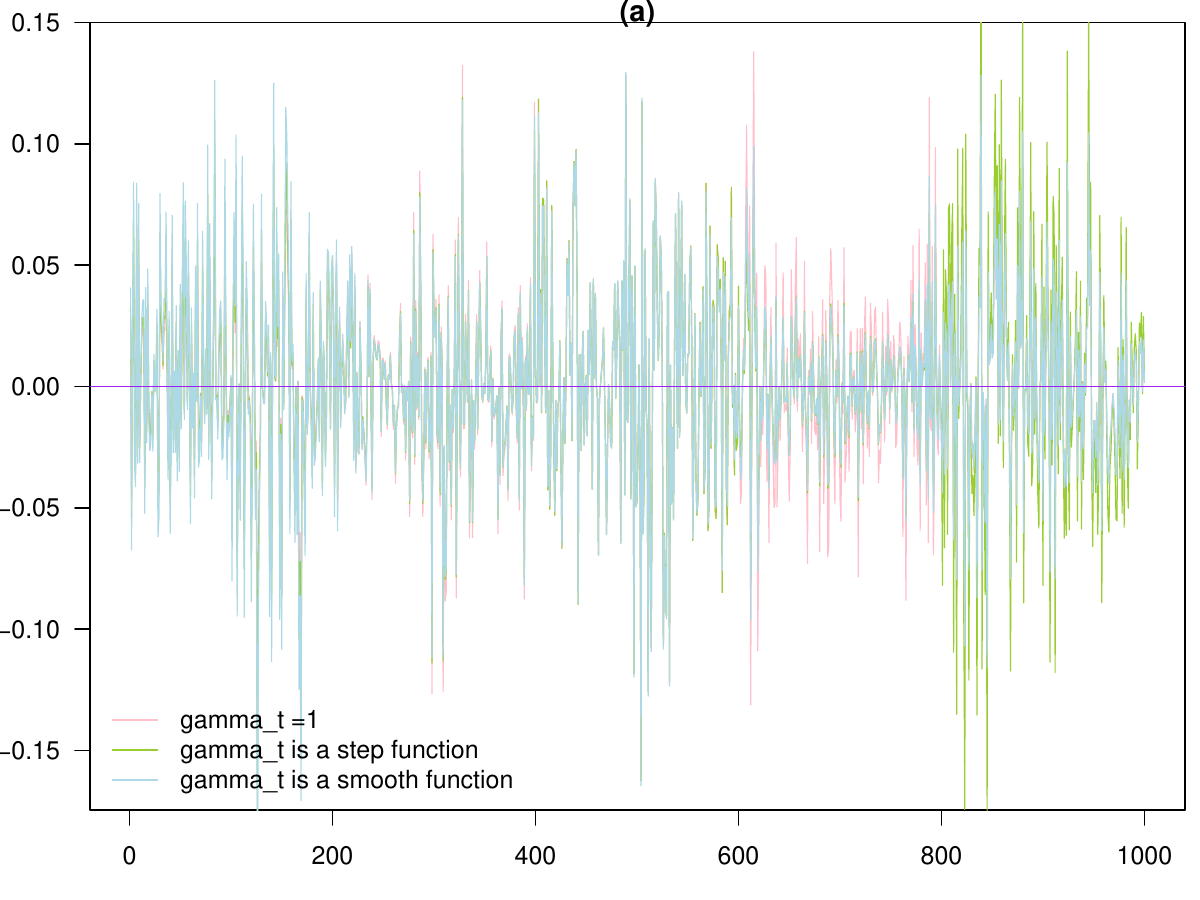}
     \caption{ Plot of time series when $\gamma_t=1$ (red), $\gamma_t$ is a step continuous function (green) and $\gamma_t$ is a smooth function (blue).}\label{fig3}
\end{figure}

This study focuses on one-step ahead prediction for VaR and ES, where the estimation for each model is done in a rolling window fashion with the window size 250 (roughly one year of data). As volatility changes rapidly, the longer window might be irrelevant in predicting future losses. The performance metrics used are root mean squared error (RMSE) and various commonly used backtesting tests where these metrics are calculated during the prediction period for each replication under each scenario.

\subsection{Results}

In this section, we present the results in terms of VaR and ES estimation with $\tau = .05$ and $.01$ under three scenarios previously described. Additionally, we incorporate the "oracle" results, where the true eGARCH models are fitted to the simulated data.

\begin{figure}[H]
    \centering
    
   \includegraphics[width=3.4in]{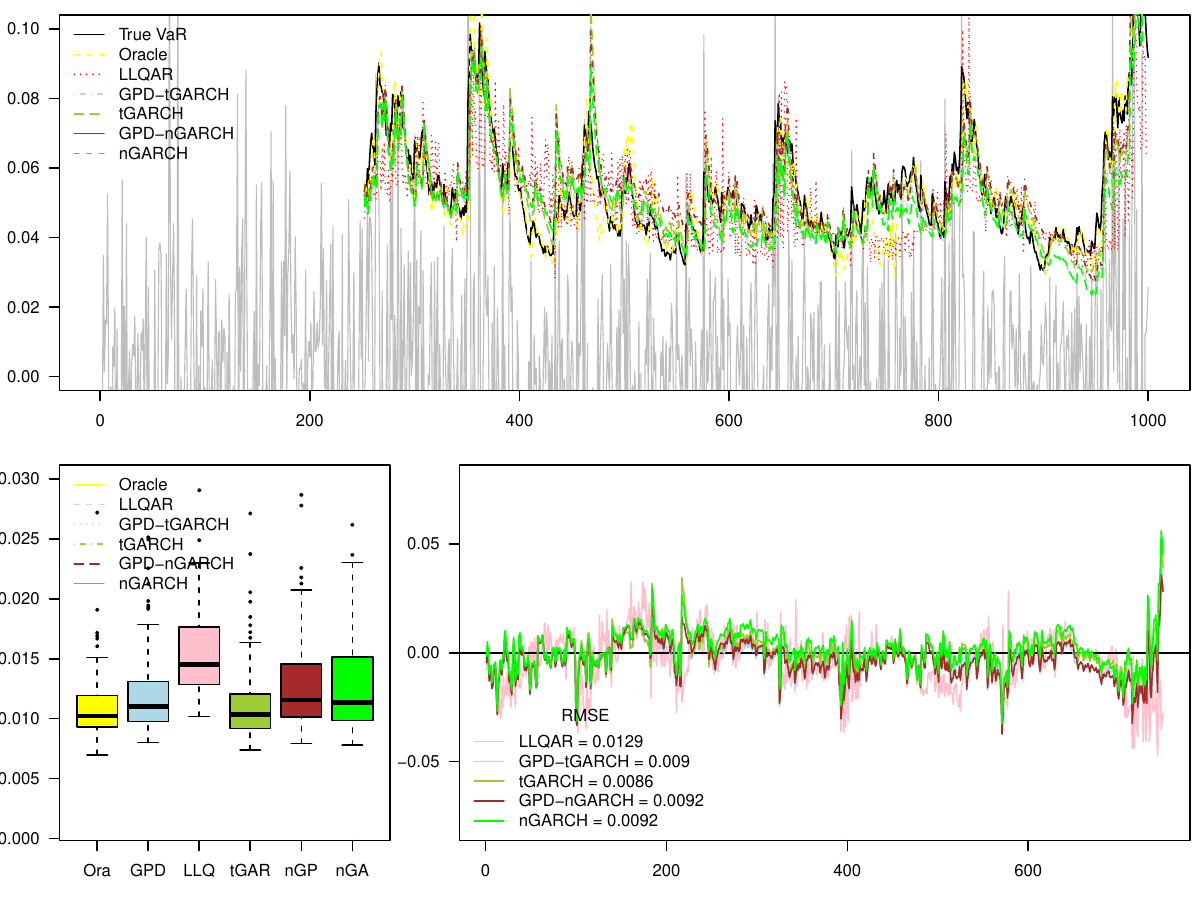}
     \caption{\textbf {Top:} One step ahead 95\% VaR predictions from a simulated data. \textbf{Bottom:} Box plot of RMSE (left) and corresponding error (right) plot of simulated data with  RMSE (Stationary dataset).}\label{fig4}
\end{figure}

\begin{figure}[H]
       \centering
       \includegraphics[width=3.4in]{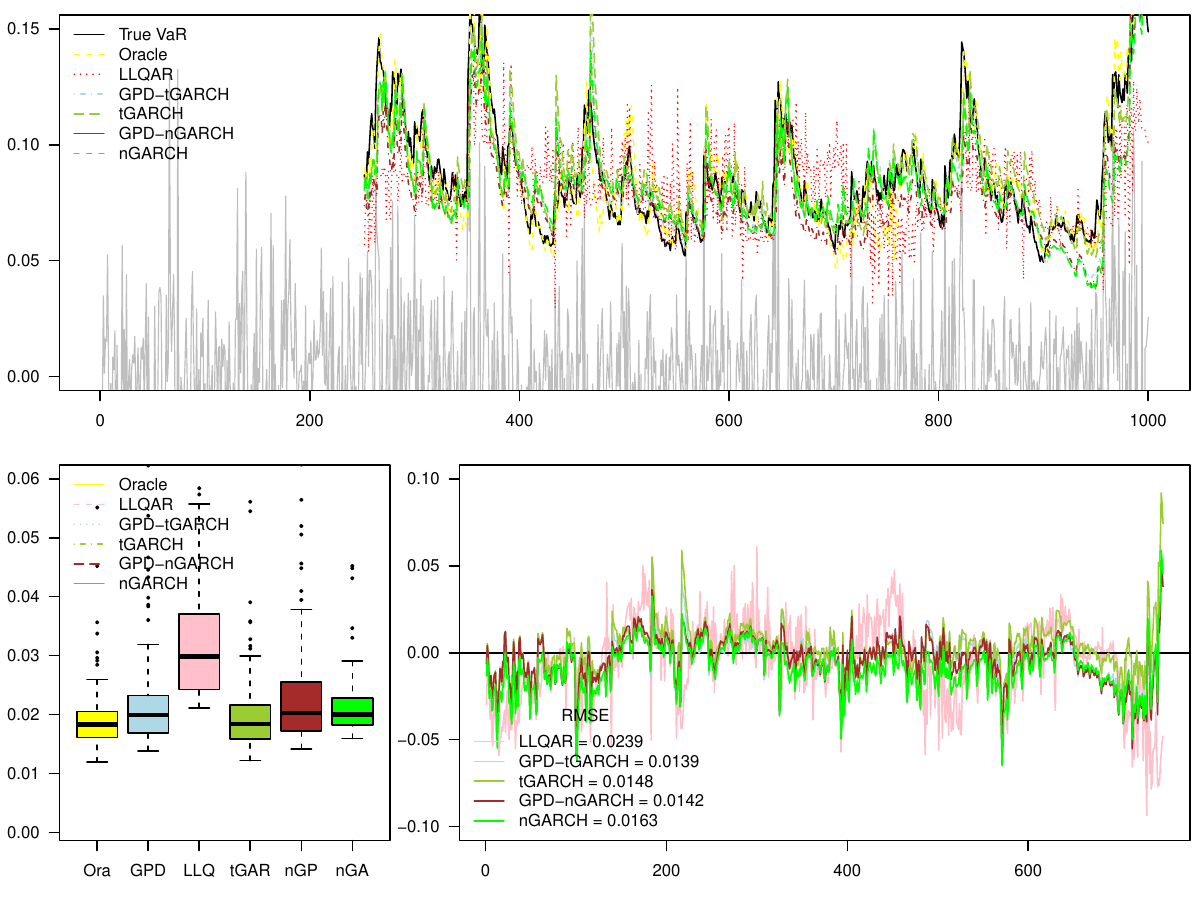}
       \caption{\textbf {Top:} One step ahead 99\% VaR forecast. \textbf{Bottom:} Box plot of RMSE (left) and corresponding error (right) plot of simulated data with  RMSE (Stationary dataset).}\label{fig5}
\end{figure}

In scenario 1 of the simulation study (Stationary case), for each sample, we estimated $95\%$ $ (\tau = .05)$ and $99\%$ $ (\tau = .01)$ quantiles (VaR) and ES using the methods provided above. 
Figure \ref{fig4} and Figure \ref{fig5}  represent the plots of the 
95\% and 99\% Value at Risk (VaR) estimates obtained from different methods, respectively. The analysis of these plots reveals that for the 
95\% quantiles, the method with the smallest Root Mean Squared Error (RMSE) is tGARCH, and overall RMSE values indicate that GPD-tGARCH and tGARCH produce results that are closely aligned. When considering the 99\% quantile, nGARCH performs similarly to tGARCH and GPD-tGARCH, with the smallest RMSE observed for tGARCH.

\begin{figure}[H]
    \centering
    \includegraphics[width=3.4in]{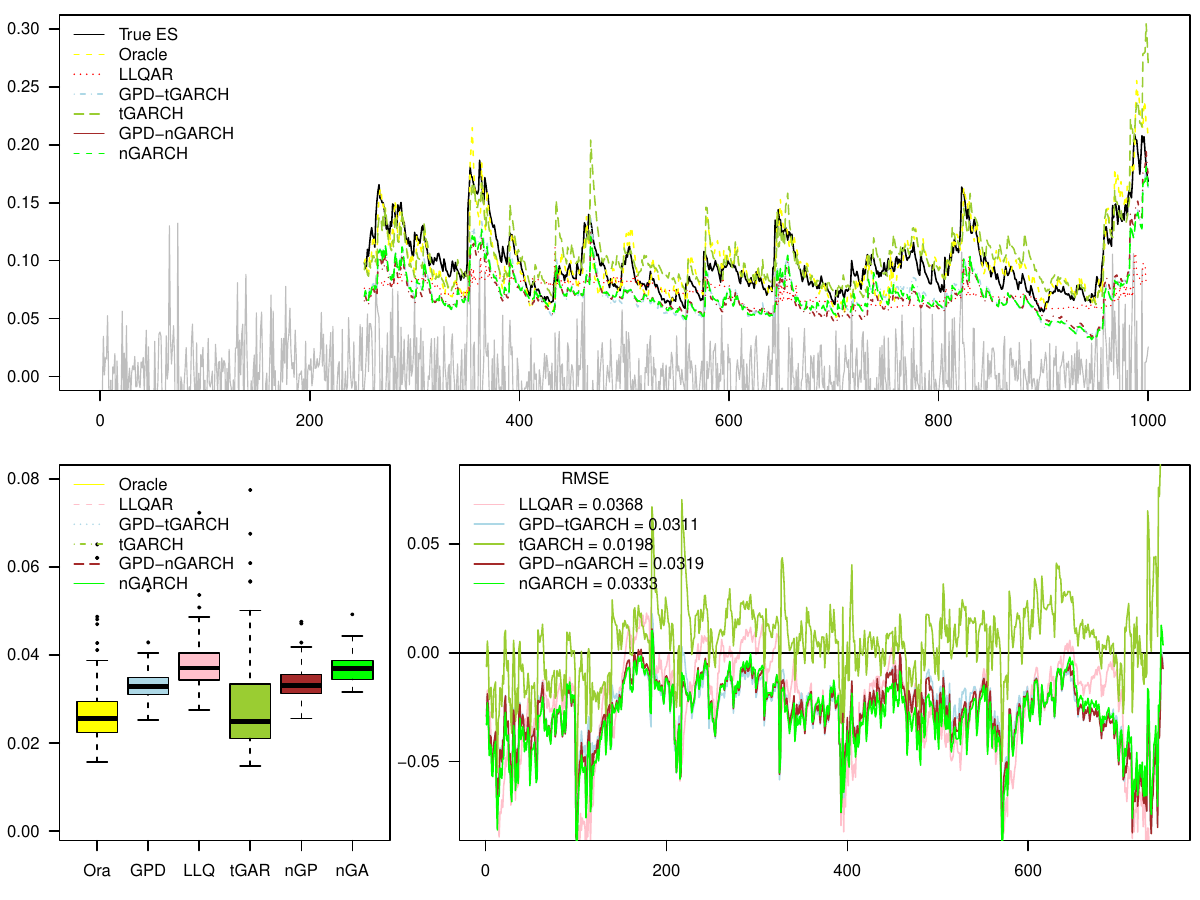}
     \caption{\textbf {Top:} One step ahead 95\% ES forecast. \textbf{Bottom:} Box plot of RMSE (left) and corresponding error (right) plot of simulated data with  RMSE (Stationary dataset).}\label{fig6}
\end{figure}

\begin{figure}[H]
       \centering
       \includegraphics[width=3.4in]{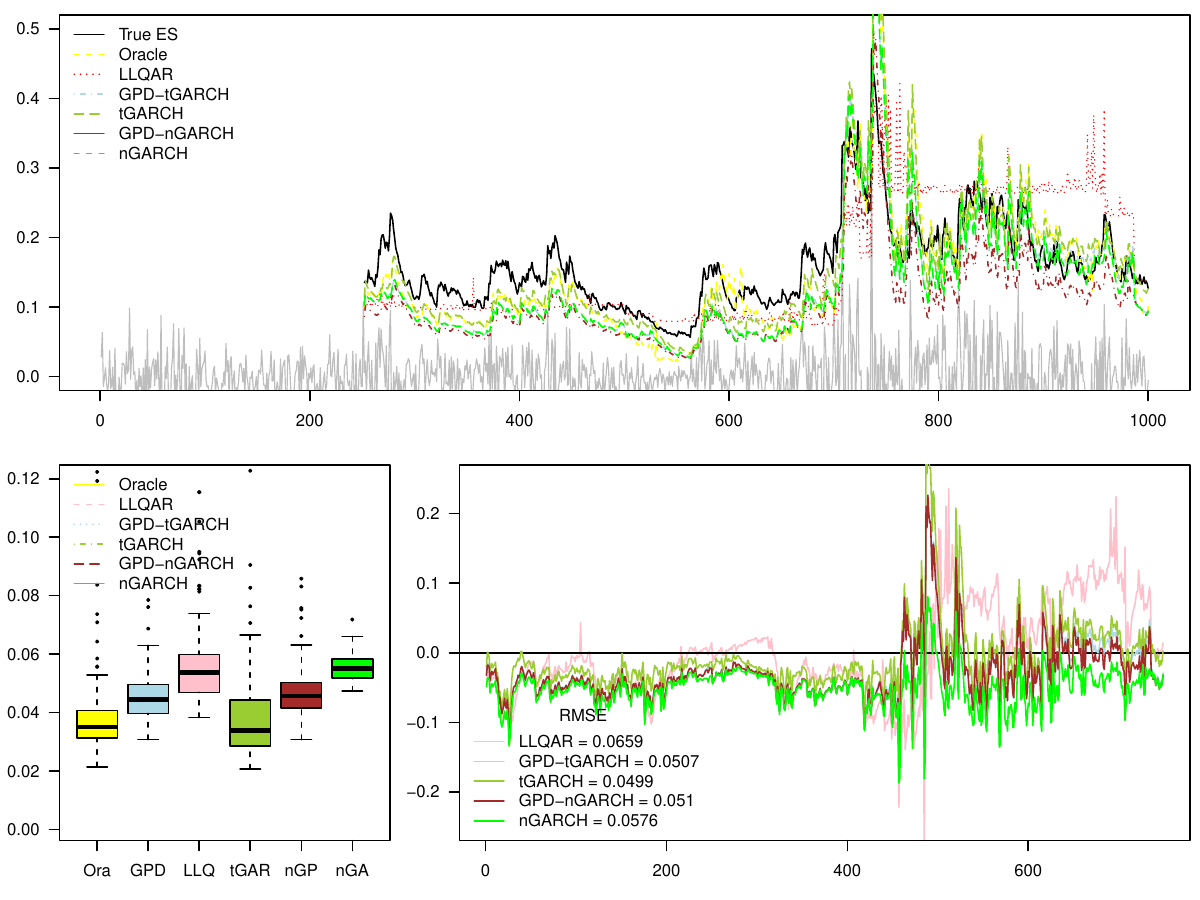}
       \caption{\textbf {Top:} One step ahead 99\% ES forecast. \textbf{Bottom:} Box plot of RMSE (left) and corresponding error (right) plot of simulated data with  RMSE (Stationary dataset).}\label{fig7}
\end{figure}

Moving on to Expected Shortfall (ES) estimation, figures \ref{fig6} and \ref{fig7} illustrate that tGARCH exhibits the best performance for both 95\% and 99\% quantiles. Throughout all cases of VaR and ES estimation, it is noted that LLQAR does not perform as well as the GARCH parametric methods.

\begin{figure}[H]
    \centering
    \includegraphics[width=3.4in]{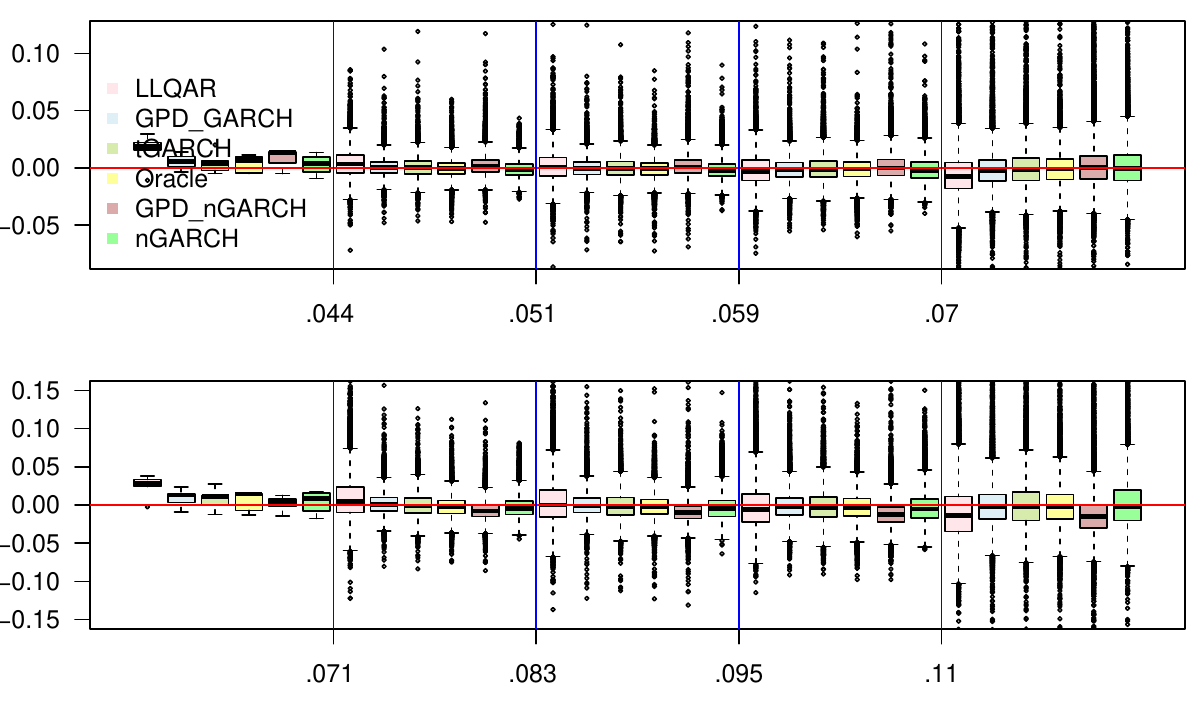}
     \caption{\textbf {Top:} Box plot of error of 95\% VaR forecast. \textbf{Bottom:} Box plot of error of 99\% VaR forecast (Stationary dataset).}\label{fig8}
\end{figure}

\begin{figure}[H]
    \centering
    \includegraphics[width=3.4in]{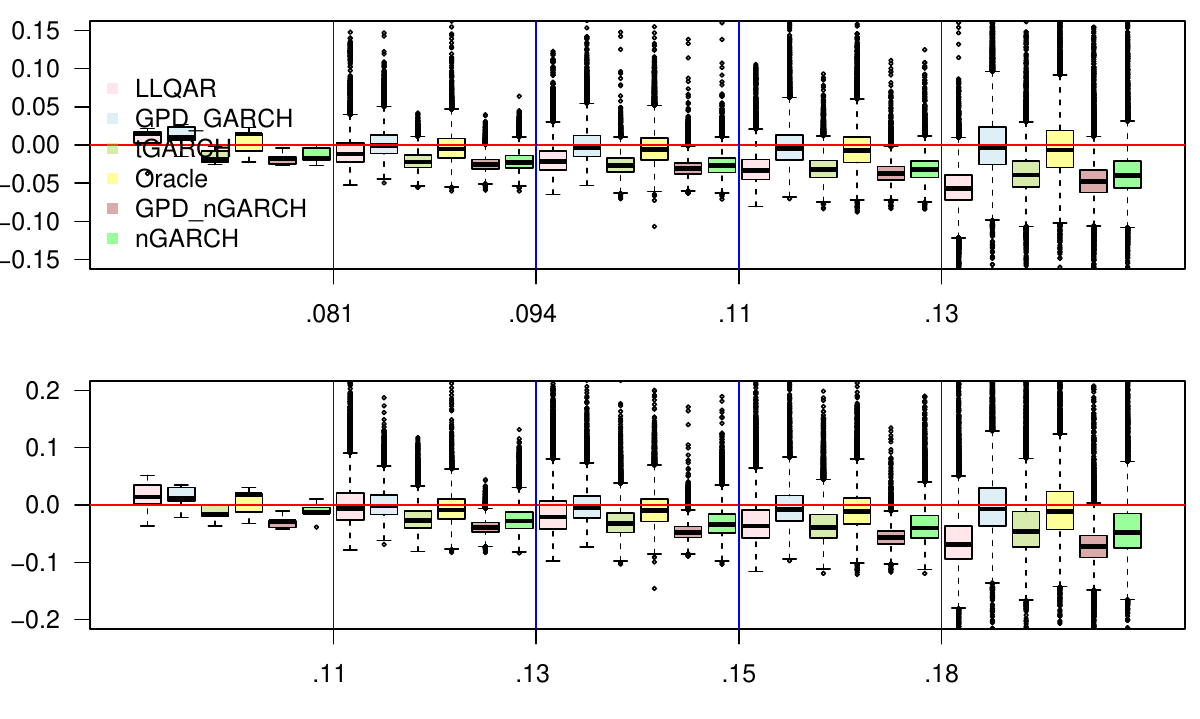}
     \caption{\textbf {Top:} Box plot of error of 95\% ES forecast. \textbf{Bottom:} Box plot of error of 99\% ES forecast (Stationary dataset).}\label{fig9}
\end{figure}

To gauge the accuracy of these estimators, an examination of their bias and variance is undertaken. The graphical results provided in Figures \ref{fig8} and \ref{fig9} offer a visual depiction of the bias and variance characteristics of the VaR estimators.
 
To gain a clearer understanding, the total risk region is segmented into five regions based on the quantiles of true Value at Risk (VaR) and true Expected Shortfall (ES). Examining the errors of VaR estimates at both quantile levels reveals noteworthy patterns.

For LLQAR, there is a tendency to overestimate in the first region, leading to high bias and variability. In contrast, GARCH family methods consistently provide more uniform estimates of risk for both quantiles, indicating superior performance compared to LLQAR.

In the middle three regions, all methods demonstrate moderate bias and variance, suggesting good overall performance. However, in higher-risk regions for both quantiles, LLQAR tends to underestimate the risk. While this signifies low bias, the high variance implies greater variability in the estimates. On the other hand, GARCH family methods generally exhibit better performance in these higher-risk regions, displaying both low bias and variance. The exception is GPD-nGARCH, which tends to underestimate.

This analysis provides a nuanced understanding of how different methods perform across different risk levels, shedding light on their strengths and weaknesses in capturing extreme losses and tail behavior.

In the investigation of the error plot for Expected Shortfall (ES) estimates, certain patterns emerge for both the 95\% and 99\%
 ES estimates across different risk regions.
In the first risk region, it is observed that both LLQAR and GPD-tGARCH tend to overreact in capturing the tail, resulting in an overestimate of the risk. In contrast, other GARCH methods tend to underestimate the risk in this region.

For all other risk regions and for both quantiles, GPD-tGARCH and the Oracle methods stand out as being able to estimate the risk more accurately. However, tGARCH, nGARCH, and GPD-nGARCH consistently tend to underestimate the risk, which is not surprising given that ES is a tail estimation method. Notably, nGARCH is found to be inadequate in capturing the extreme tail behavior, resulting in underestimation.

The variability in estimates is highest for GPD-tGARCH and Oracle in the higher-risk regions for both quantiles, indicating a more volatile estimation process. On the other hand, LLQAR provides accurate estimation for the second risk region of both quantiles but tends to underestimate in all other cases.

This detailed analysis provides insights into the performance of different methods in capturing tail behavior and estimating Expected Shortfall across various risk levels.

In summary, within scenario 1, it is evident that for both quantiles of Value at Risk (VaR) estimation, tGARCH stands out as the best performer. Additionally, GPD-tGARCH demonstrates a performance that is closely aligned with tGARCH for VaR estimation.

For Expected Shortfall (ES) estimation, tGARCH exhibits the smallest Root Mean Squared Error (RMSE), yet it consistently underestimates the risk. On the other hand, GPD-tGARCH provides unbiased estimates in both quantiles, positioning it as the best performer for ES estimation.

In scenario 2, characterized by a non-stationary case where $\gamma_t$ is selected as a continuous step function, introducing non-stationarity with notable abrupt fluctuations in volatility, the analysis focuses on the 95\% and 
99\% Value at Risk (VaR) estimates (Figures 10 and 11) and Expected Shortfall (ES) estimates (Figures 12 and 13) for both quantiles. These figures provide a visual representation of how different methods perform in capturing extreme losses and tail behavior under non-stationary conditions. 

\begin{figure}[H]
    \centering
    
   \includegraphics[width=3.4in]{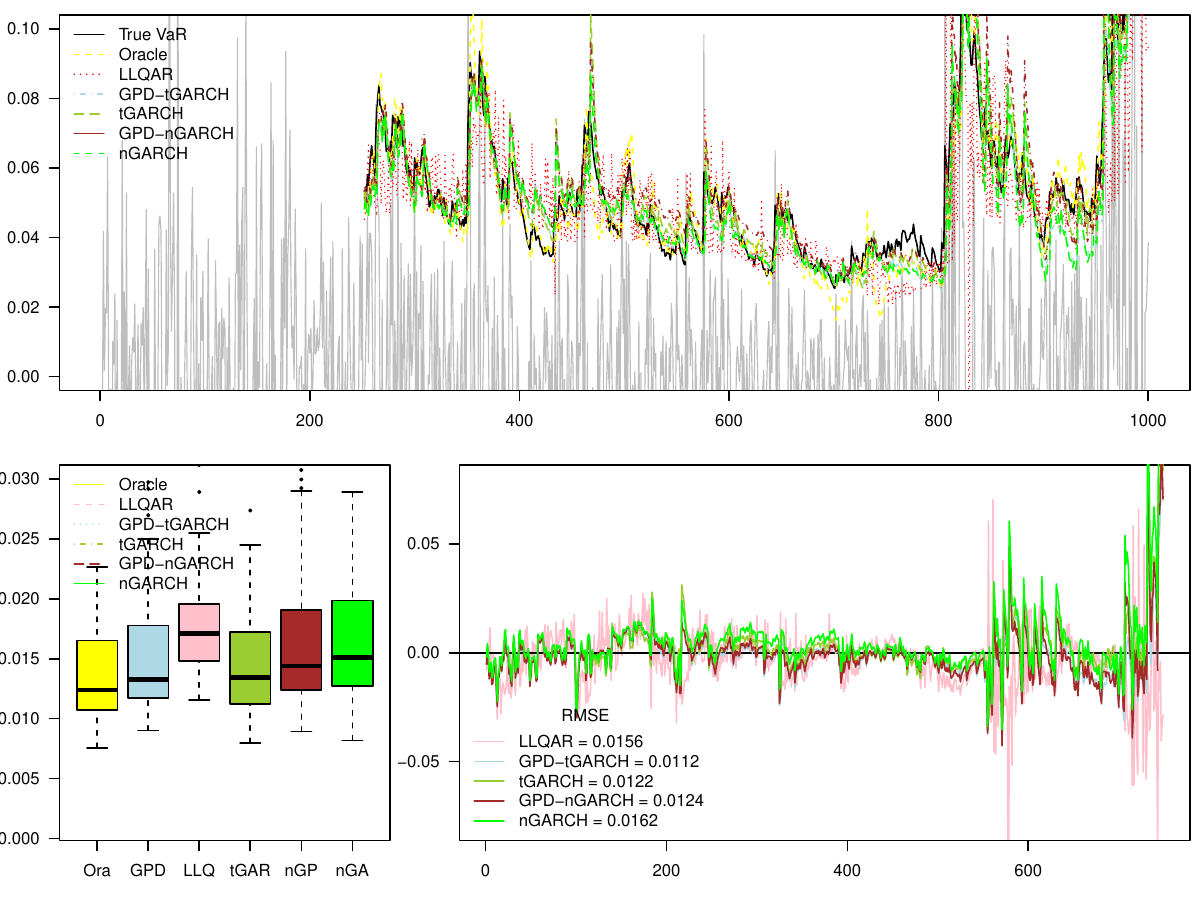}
     \caption{\textbf {Top:} One step ahead 95\% VaR predictions from a simulated data. \textbf{Bottom:} Box plot of RMSE (left) and corresponding error (right) plot of simulated data with  RMSE (Non-Stationary dataset).}\label{fig10}
\end{figure}

\begin{figure}[H]
       \centering
       \includegraphics[width=3.4in]{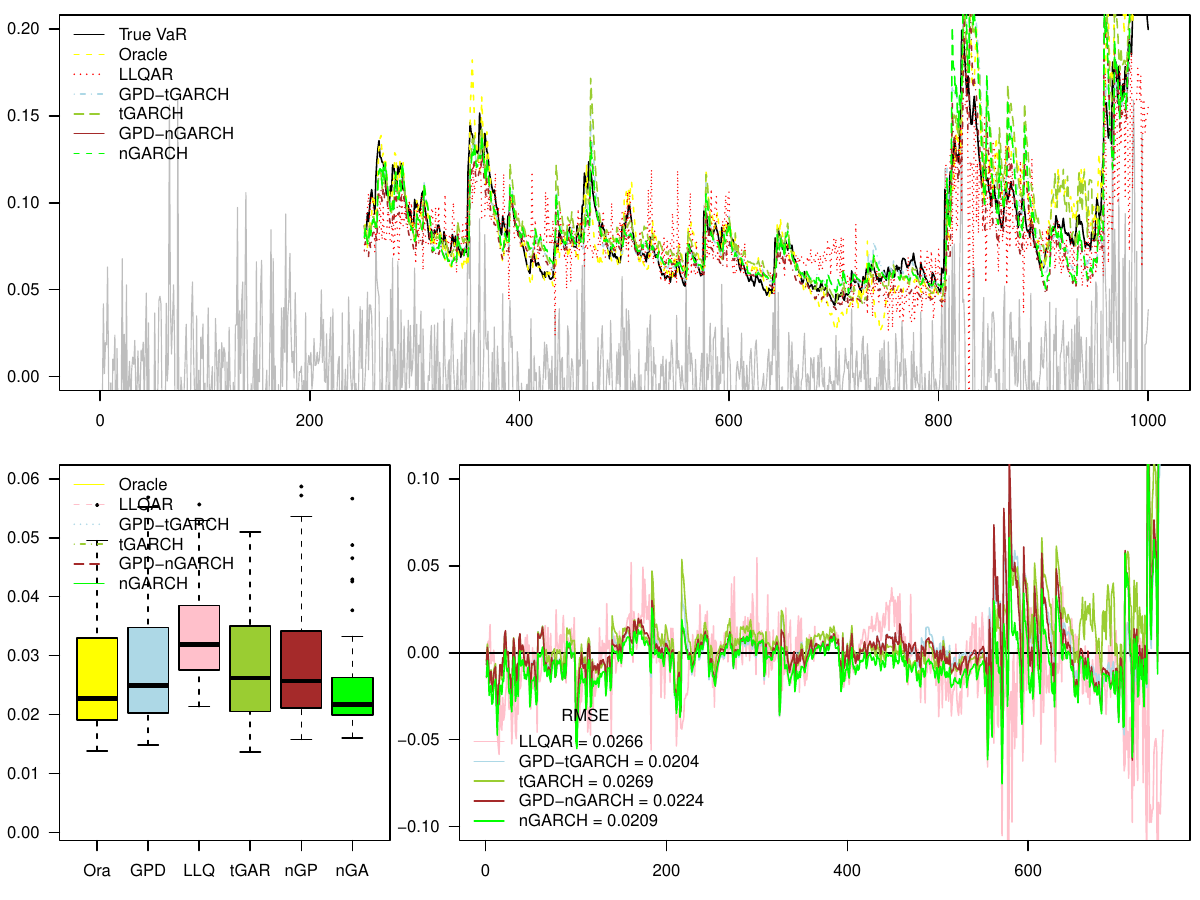}
       \caption{\textbf {Top:} One step ahead 99\% VaR forecast. \textbf{Bottom:} Box plot of RMSE (left) and corresponding error (right) plot of simulated data with  RMSE (Non-Stationary data set).}\label{fig11}
\end{figure}

In scenario 2, the performance of the Local Linear Quantile Autoregression (LLQAR) method demonstrates improvement compared to scenario 1, particularly for Value at Risk (VaR) estimations at both quantiles. LLQAR competes favorably with GARCH family methods in VaR estimations, indicating its efficacy under non-stationary conditions. However, it is noted that if there are several shocks close to each other, the proposed method may recover slowly compared to other benchmark methods for both quantiles. Nevertheless, LLQAR responds quickly in scenarios without closely spaced shocks, similar to benchmark methods.
For 95\% VaR estimation in figure \ref{fig10}, tGARCH stands out with the lowest Root Mean Squared Error (RMSE). For 
99\% quantile VaR estimation in figure \ref{fig11}, all GARCH models perform similarly well, with nGARCH having the smallest RMSE.

\begin{figure}[H]
    \centering
    \includegraphics[width=3.4in]{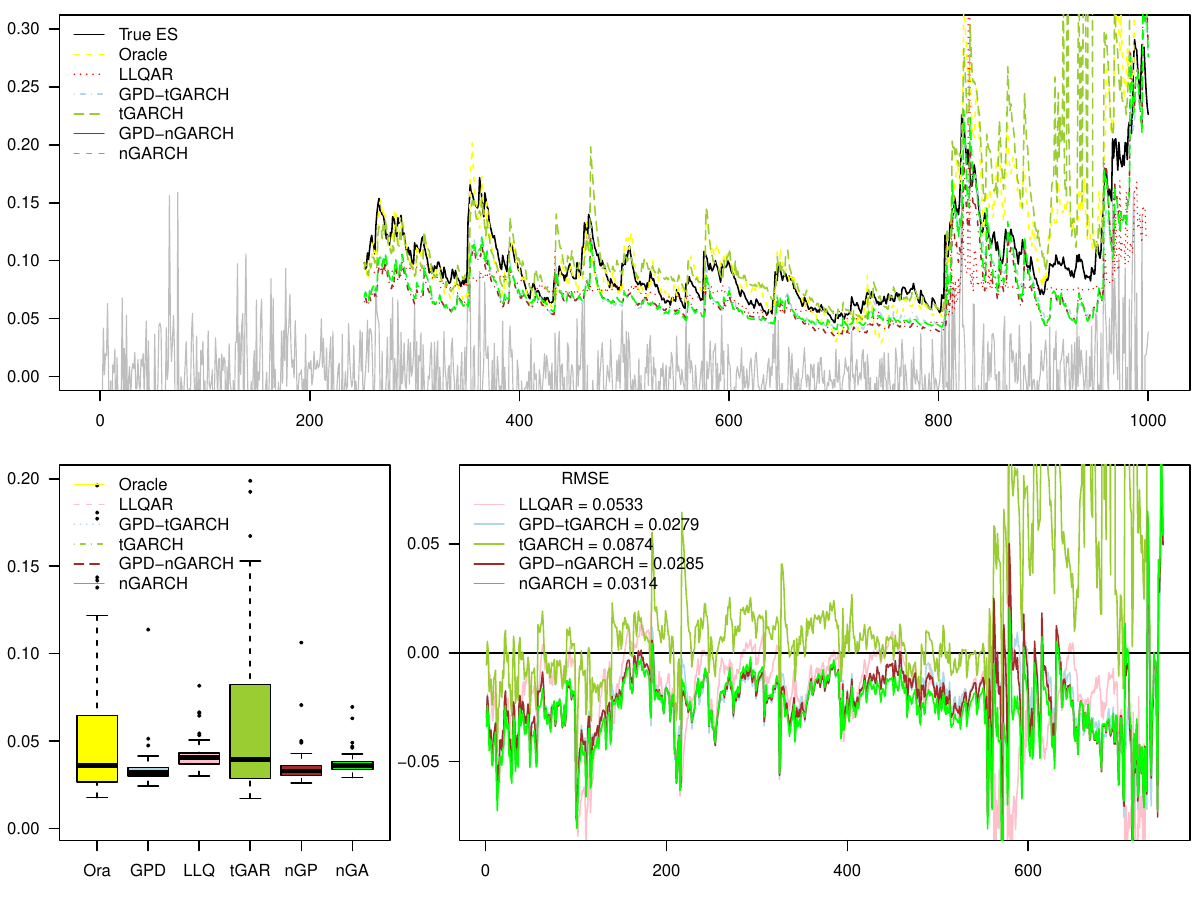}
     \caption{\textbf {Top:} One step ahead 95\% ES forecast. \textbf{Bottom:} Box plot of RMSE (left) and corresponding error (right) plot of simulated data with  RMSE (Non-Stationary data set).}\label{fig12}
\end{figure}

\begin{figure}[H]
       \centering
       \includegraphics[width=3.4in]{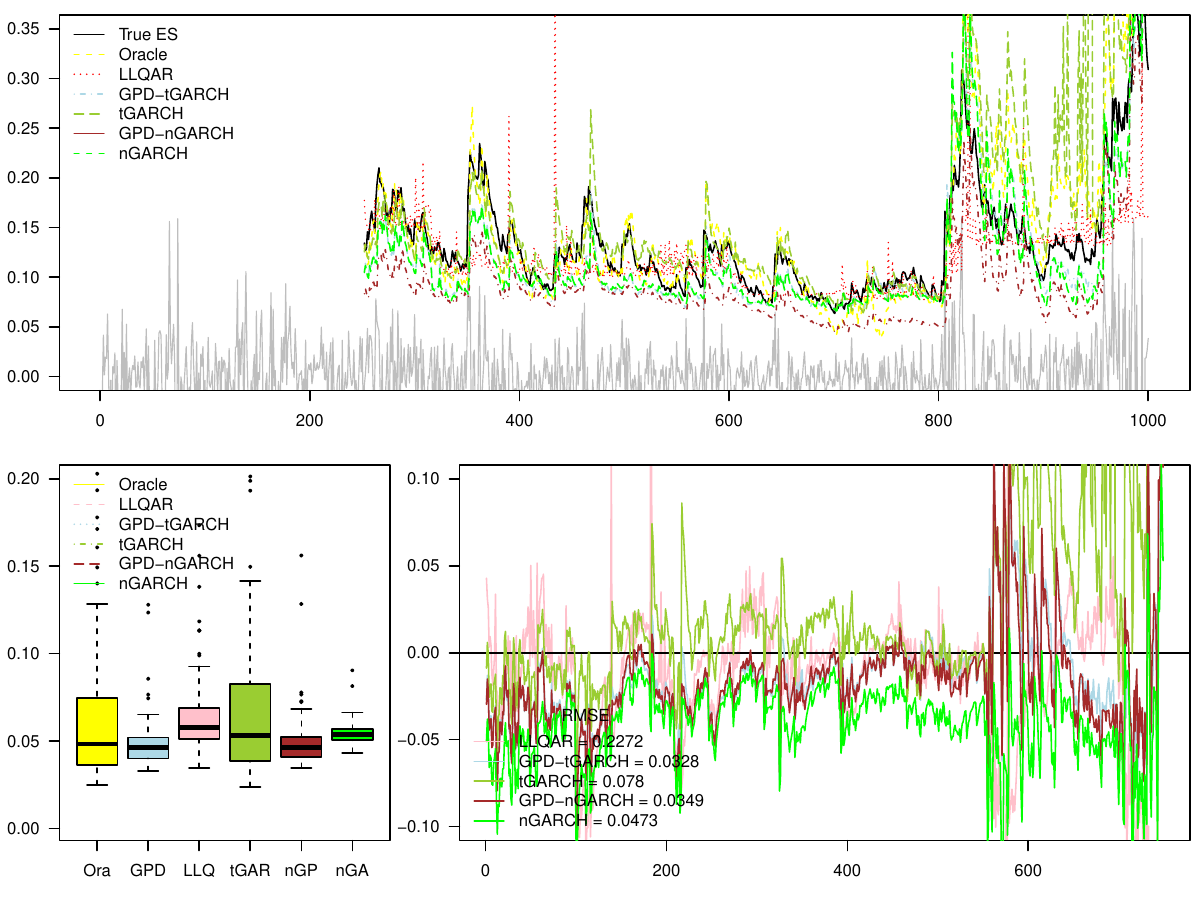}
       \caption{\textbf {Top:} One step ahead 99\% ES forecast. \textbf{Bottom:} Box plot of RMSE (left) and corresponding error (right) plot of simulated data with  RMSE (Non-Stationary data set).}\label{fig13}
\end{figure}

In terms of Expected Shortfall (ES) estimation figures \ref{fig12} and \ref{fig13}, LLQAR demonstrates improved performance. GPD-tGARCH and GPD-nGARCH exhibit the smallest RMSE, while LLQAR provides the second lowest RMSE but with the highest observed variability, particularly in comparison to tGARCH. Notably, both nGARCH and GPD-nGARCH have a tendency to underestimate risk frequently.

Overall, LLQAR shows enhanced performance in scenario 2, competing well with GARCH family methods for VaR estimations, and its performance in ES estimation also sees improvement. The observations highlight the nuanced behavior of different methods in capturing risk under non-stationary conditions.

Analyzing the error plots of both Value at Risk (VaR) and Expected Shortfall (ES) provides insights into the performance of different methods under non-stationary conditions.

\begin{figure}[H]
    \centering
    \includegraphics[width=3.4in]{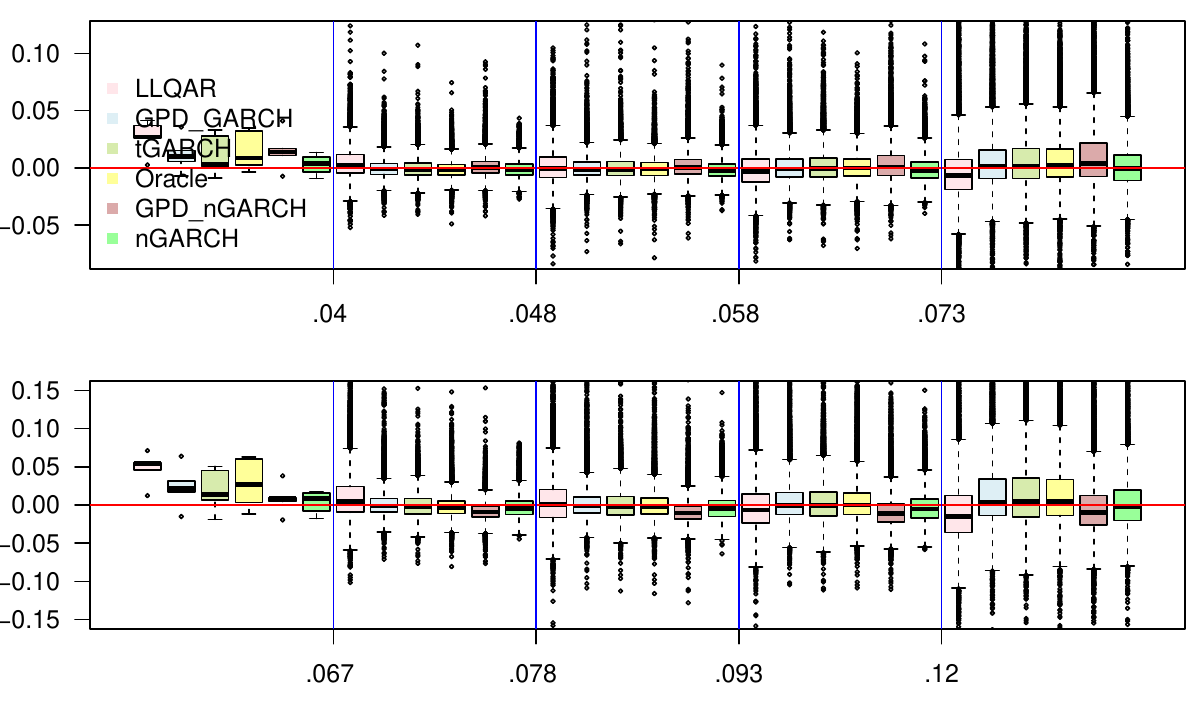}
     \caption{\textbf {Top:} Box plot of error of 95\% VaR forecast. \textbf{Bottom:} Box plot of error of 99\% VaR forecast (Non-Stationary dataset).}\label{fig14}
\end{figure}

For VaR estimation figure \ref{fig14}, in the first risk region, all methods except nGARCH tend to overestimate, with LLQAR exhibiting the highest bias for both quantiles. In the middle three risk regions of VaR for both quantiles, all methods perform similarly well, generating unbiased estimates. In the high-risk region, LLQAR tends to underestimate with higher variability compared to other methods for both quantiles.

\begin{figure}[H]
    \centering
    \includegraphics[width=3.4in]{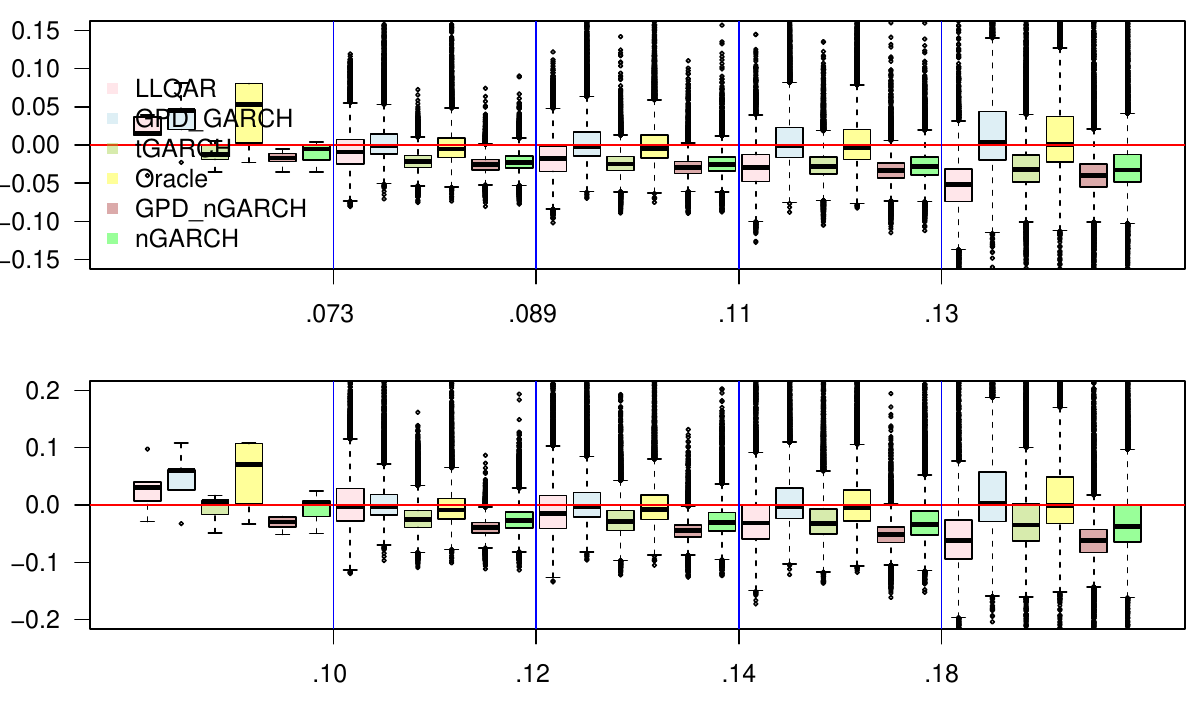}
     \caption{\textbf {Top:} Box plot of error of 95\% ES forecast. \textbf{Bottom:} Box plot of error of 99\% ES forecast (Non-Stationary dataset).}\label{fig15}
\end{figure}

For ES estimation figure \ref{fig15}, LLQAR, GPD-tGARCH, and Oracle tend to overestimate with high variability observed in Oracle, while other methods tend to underestimate with small variability in the first risk region. Across all regions and both quantiles of ES, tGARCH, nGARCH, and GPD-nGARCH consistently underestimate risk, whereas other methods provide more robust estimates. Only in high-risk regions of both quantiles does LLQAR tend to underestimate. Among all methods, GPD-tGARCH and Oracle exhibit more robust and unbiased performance for ES estimation.

In summary, it is concluded that tGARCH performs the best in 95\% VaR estimates, nGARCH is the best for 99\% VaR, and GPD-tGARCH is the best performer for both quantiles in ES estimation. LLQAR demonstrates the second-best performance in both VaR and ES estimation. These findings highlight the nuanced strengths and weaknesses of different methods under non-stationary conditions.

It appears that the results for Scenario 3, described as close to non-stationary, exhibit a similar pattern to Scenario 1. Due to the similarity in patterns and to avoid redundancy, the details of the results for Scenario 3 are left unexplained in the main text and are presented in the Appendix.

The summary of test rejections for all methods provides additional insights into the performance of these methods in different scenarios are  displayed in table 1.

For the Unconditional Coverage (UC) test, which assesses the correct number of exceedances, all methods demonstrate a percentage of test rejections close to the significance level, except for LLQAR. LLQAR tends to have a slightly higher percentage of test rejections across all scenarios for the 95\% Value at Risk (VaR) estimate. However, for the 99\% VaR case, both LLQAR and nGARCH exhibit a higher percentage of test rejections. The elevated percentage for nGARCH is expected due to the tailed nature of the data.

Regarding the Conditional Coverage (CC) test, which evaluates correct exceedances and independence, all methods show a higher proportion of rejections than the significance level. This suggests that the violations are clustered for both quantiles.

In the Expected Shortfall (ES) bootstrap test, tGARCH and LLQAR display a percentage of test rejections that are closest to the significance level for both quantiles. In contrast, other methods exhibit a higher percentage of test rejections.

These test results further contribute to the comprehensive evaluation of the methods, offering insights into their ability to correctly capture exceedances, independence, and clustering behavior under different scenarios.


The overall conclusion is that different methods exhibit better performance under different datasets and scenarios. In stationary cases, the top-performing models for Value at Risk (VaR) are tGARCH and GPD-tGARCH. In non-stationary cases, tGARCH and nGARCH models perform well, particularly when volatility experiences sudden and significant shocks. LLQAR provides competitive estimates for VaR estimation in non-stationary cases, while GPD-tGARCH emerges as the best-performing model for Expected Shortfall (ES) estimation in both quantiles.

It's noteworthy that no single method consistently outperforms others across all datasets and scenarios. The effectiveness of each method depends on the specific characteristics and dynamics of the data, emphasizing the importance of selecting an appropriate model based on the nature of the dataset and the underlying phenomena.


The observation that GARCH family models fail to converge for certain periods of time series due to non-stationarity, while LLQAR can converge in every period and every case, is a noteworthy finding. It underscores a limitation of GARCH models, especially in handling highly non-stationary data with abrupt changes in volatility. GARCH models are designed under the assumption of stationarity, and when faced with non-stationary conditions, they may struggle to converge and provide reliable estimates.

On the other hand, the robustness of LLQAR in converging across various periods and cases, irrespective of the non-stationarity, highlights its flexibility and adaptability to different data structures. This characteristic makes LLQAR a potentially valuable tool, particularly in scenarios where traditional parametric models face challenges due to non-stationarity.

\begin{table}
\begin{center}
 \scalebox{.85}{
    \begin{tabular}{p{4cm}|c|c|c|c|c|c|c}\hline
 Simulations & Methods & \multicolumn{2}{c|}{UC}  & \multicolumn{2}{c|} {CC}  &\multicolumn{2}{c} {ES Bootstrap}\\
\cline{3-8}
   & & $\tau=.05$& $\tau=.01$ &  $\tau=.05$& $\tau=.01$& $\tau=.05$& $\tau=.01$\\
    \hline
    
 
 & tGARCH & \textcolor{red} {10\%}&\textbf{6}\%&\textcolor{red} {12\%}&\textcolor{red} {27\%}&\textcolor{red} {0}&\textbf{1} \\

 & nGARCH & \textbf{5}\%&\textcolor{red} {68\%}&\textcolor{red} {24\%}&\textcolor{red} {61\%}&\textcolor{red} {67\%}&\textcolor{red} {64\%} \\
 
Simulation Study-1 (out of 100)&GPD-tGARCH & \textbf{2\%}&\textbf{9\%}&\textcolor{red} {23\%}&\textcolor{red} {19\%}&\textcolor{red} {14\%}&\textcolor{red} {15\%}\\

&GPD-nGARCH & \textbf{3\%}&\textcolor{red} {12\%}&\textcolor{red} {26\%}&\textcolor{red} {20\%}&\textcolor{red} {11\%}&\textcolor{red} {22\%}\\

& LLQAR & \textcolor{red} {12\%}&\textcolor{red} {25\%}&\textcolor{red} {13\%}&\textcolor{red} {21\%}&\textbf{4}\%&\textbf{1\%}\\\hline


& tGARCH & \textbf{8.24\%} & \textcolor{red} {11.3\%} &\textcolor{red} {25.7\%} &\textcolor{red} {17.5\%} &\textcolor{red} {0}& \textcolor{red} {0}\\

& nGARCH & \textbf{8.24\%} & \textcolor{red} {60.8\%} &\textcolor{red} {26.88\% }&\textcolor{red} {58.7\%} &\textcolor{red} {78.35}& \textcolor{red} {62.8\%}\\
 
Simulation Study-2 (Out of 97)& GPD-tGARCH & \textbf{6.18\%}   &\textcolor{red} {11.3\% }& \textcolor{red} {23.7\%}&\textcolor{red} {21.6\%} & \textbf{5.15\%}&\textcolor{red} {18.5\%}\\

&GPD-nGARCH & \textbf{4.12\% }  &\textcolor{red} {13.4\%} & \textcolor{red} {21.6\%}&\textcolor{red} {18.5\%} & \textcolor{red} {14.4\%}&\textcolor{red} {19.5\%}\\

& LLQAR & \textcolor{red} {19.6\%}&\textcolor{red} {42.2\%} &\textcolor{red} {18.6\%}&\textcolor{red} {37.1\%} & \textbf{7.2\%}&\textcolor{red} {0\%}\\\hline


& tGARCH & \textbf{7.14\%} & \textcolor{red} {9.18\%} &\textcolor{red} {27.5\%} &\textbf{4.08\%} &\textcolor{red} {0}& \textbf{1.02}\\

& nGARCH & \textbf{5.10\% }& \textcolor{red} {51.02\%} &\textcolor{red} {23.5\%} &\textcolor{red} {51.02\%} &\textcolor{red} {78.6}& \textcolor{red} {60.2\%}\\
 
Simulation Study-3 (Out of 98)& GPD-tGARCH & \textbf{3.06\%}   &\textbf{6.12\%} & \textcolor{red} {23.4\%}&\textcolor{red} {16.3\%} & \textcolor{red} {10.2\%}&\textcolor{red} {17.3\%}\\

&GPD-nGARCH & \textbf{4.08\%}   &\textcolor{red} {11.2\%} & \textcolor{red} {19.3\%}&\textcolor{red} {17.3\%} & \textcolor{red} {11.2\%}&\textcolor{red} {18.3\%}\\

& LLQAR &\textcolor{red} { 9.18\%}&\textcolor{red} {26.5\% }&\textcolor{red} {12.2\%}&\textcolor{red} {23.4\%} & \textbf{3.06\%}&\textbf{3.06\%}\\\hline

\end{tabular}}  
\caption{ Summary of backtest results for each $\tau$ level for simulations. Percentage of test rejections are summarized here for each methods. Positive outcomes are highlighted in bold, while negative outcomes (two standard deviation away from nominal level) are indicated in red color.}
\end{center}
\end{table}


\section{Application}

The estimation methods of VaR and ES are applied to four major stocks IBM Common Stock, Apple Stock, S\&P 500 and Walmart Stock from January 3, 2007, to November 11, 2022. We choose this time interval to notice how the above-discussed methods perform during the financial crisis which happened in 2008 and afterward. The data sets were collected from yahoo finance. The data consists of daily closing prices for each of the stocks. At first, for each data set, log losses were computed using the equation (1). Then the VaR and ES were estimated using the rolling window of size 250 to lessen the risk of structural changes and forecast the one-step-ahead VaR and ES. The window size plays an important role in the forecast estimation. But there are no advanced studies on how to choose the optimal window size. Possible research can be carried out on how to choose the optimal window size that will yield more accurate estimates.

Graphical representation of log-losses of four different stocks are presented bellow.

\begin{figure}[H]
    \centering
    \includegraphics[width=3.4in]{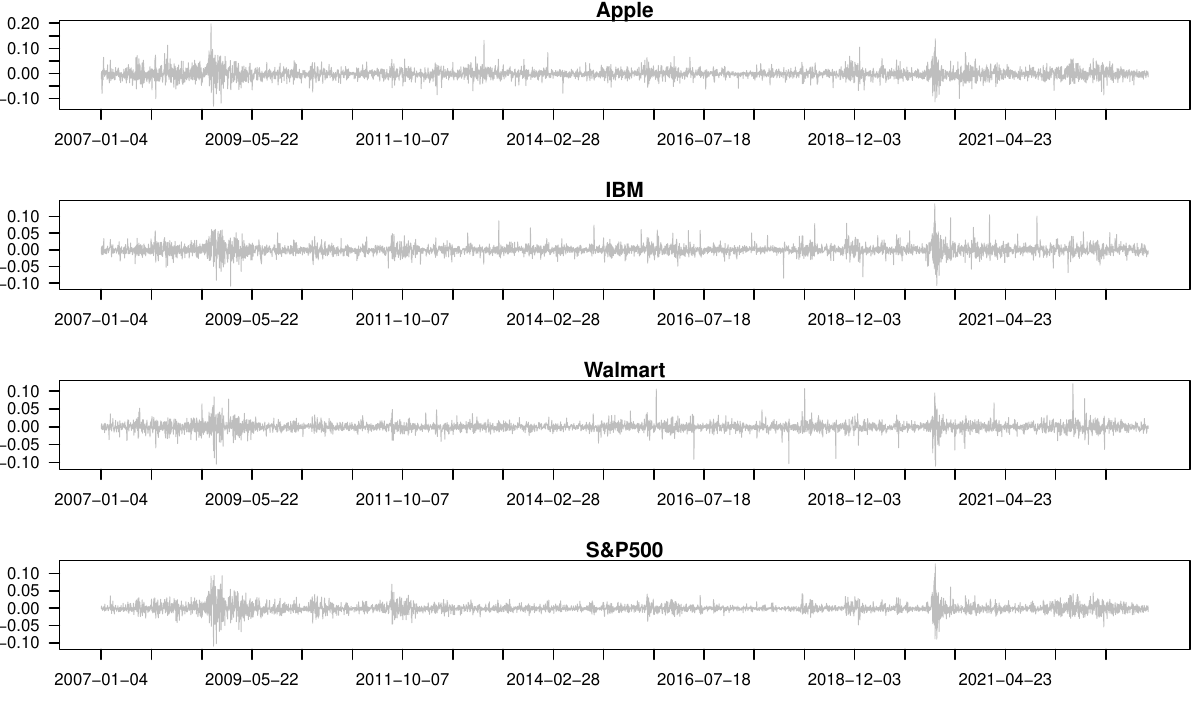}
     \caption{\tiny Time series plot of four different stocks.}
\end{figure}

\begin{figure}[H]
    \centering
    \includegraphics[width=3.4in]{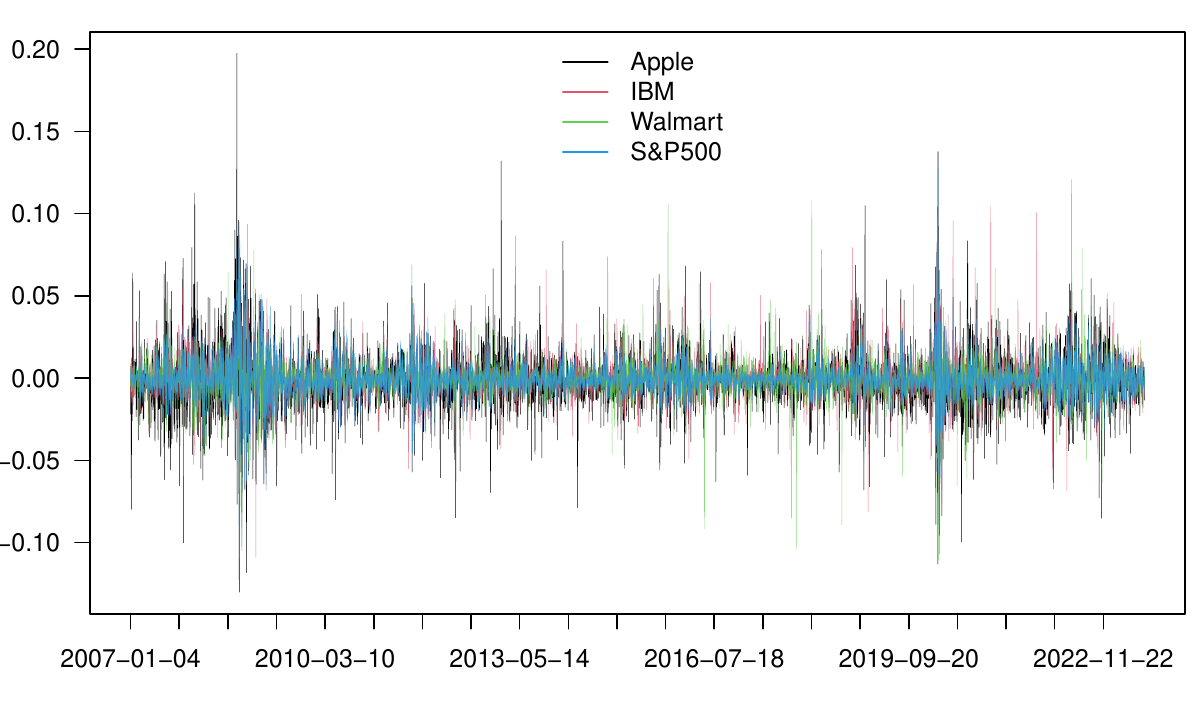}
     \caption{\tiny Combine plot of log-losses of above four stocks.}
\end{figure}

Table \ref{statistics-log-los} gives an overview of descriptive statistics of log-losses data of the stocks. The log losses for all stocks exhibit positive skewness, indicating a rightward tilt in their distribution. Additionally, all stocks display significant excess kurtosis, rejecting the assumption of normality. This suggests that the log loss distributions have heavier tails and a more pronounced peak compared to a normal distribution.

\begin{table}

    \small\begin{tabular}{l|c|c|c|c|c|c}\hline
      Stocks & Mean & SD & Skewness & Kurtosis& Min & Max\\
      \hline

 S $\&$ P 500 &-0.00025& 0.01301&0.54036&11.95642&-0.10957&0.12765\\\hline
 
 Apple Stock & -0.00099&0.02034&0.40669&6.53025&-0.13019&0.19747 \\\hline
 
IBM Common Stock &-8.975e-05&0.01513&0.43959&8.13406&-0.10899&0.13755\\\hline

Walmart Stock &-0.00027&0.01316&0.06781&12.8639&-0.11072&0.12076\\\hline

\end{tabular}
\begin{center}
\caption{Summary statistics of daily log losses of each stock.}\label{statistics-log-los}
\end{center}
\end{table}

\begin{figure}[H]
    \centering
     
       \includegraphics[width=4.5in]{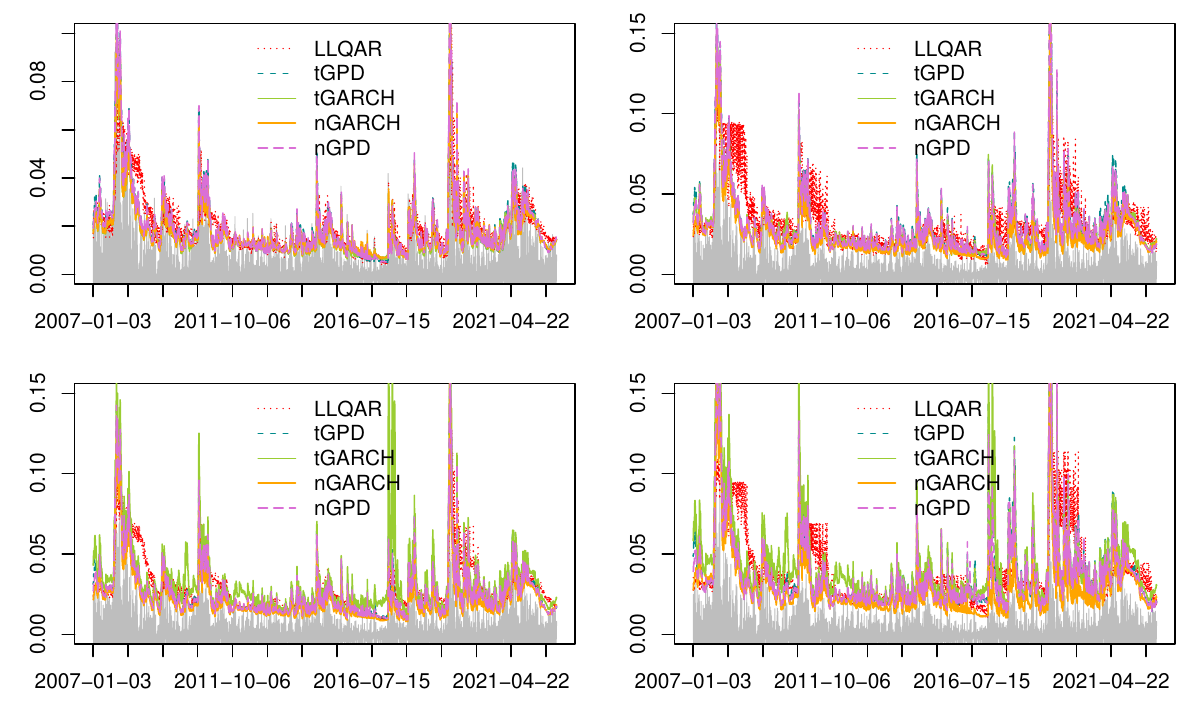}
       
      \caption{\textbf{Top}: Plot of $95\%$ (left) and $99\%$(right) VaR of different methods. \textbf{Bottom}: Plot of  $95\%$ (left) and $99\%$(right) ES of different methods for S \& P 500  Index data set.}\label{IBM-stock}
    \end{figure}

Figure \ref{IBM-stock}-\ref{S-P-500-stock} represent the VaR and ES estimated by various proposed methods of four different stocks. We observe that, across all stocks, GARCH methods demonstrate swift responses to substantial volatility changes, whereas the LLQAR method exhibits a more cautious approach, responding slowly after encountering a significant shock. The findings also indicate that, at a 95\% confidence level for Value at Risk (VaR), both GARCH methods and LLQAR offer visually comparable estimates for all stocks and the index. However, at a 99\% confidence level, GARCH methods consistently produce stable estimates, while LLQAR estimates tend to be more volatile and have a tendency to overestimate. This trend is consistent across all datasets analyzed in this study. Additionally, the outcomes demonstrate that, for Expected Shortfall (ES), tGARCH effectively captures all extreme events, albeit with a tendency to overestimate ES for both quantiles across all datasets. In contrast, GPD-tGARCH and LLQAR provide smoother and more stable estimates but occasionally underestimate ES. The comparison also extends to parametric, semi-parametric, and non-parametric methods against CAViaR models. Notably, CAViaR exhibits inferior performance compared to LLQAR. Further information on these methods is available in \cite{engle2004caviar}. The graphical comparison of these methodologies can be accessed in Appendix C.\\
\\


    \begin{figure}[H]
    \centering
      \includegraphics[width=4.5in]{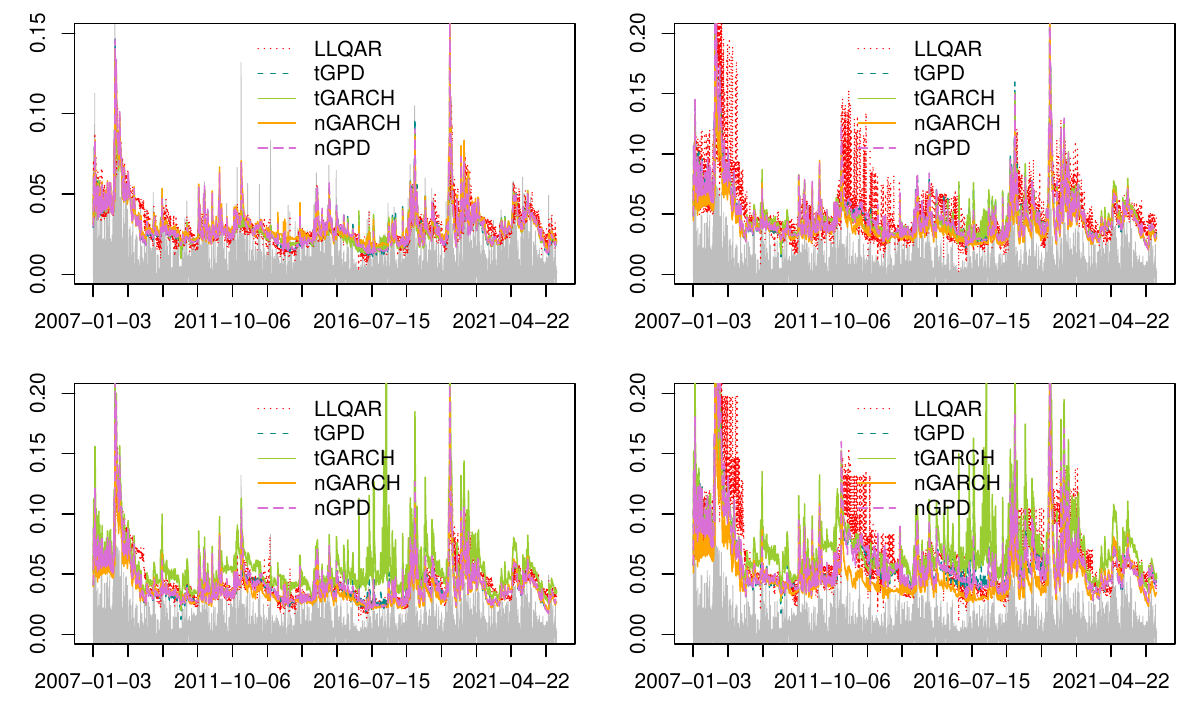}
       
      \caption{\textbf{Top}: Plot of $95\%$ (left) and $99\%$(right) VaR of different methods. \textbf{Bottom}: Plot of  $95\%$ (left) and $99\%$(right) ES of different methods for Apple Stock  data set.}\label{Apple-stock}
    
    \end{figure}

    \begin{figure}[H]
    \centering
     
       \includegraphics[width=4.5in]{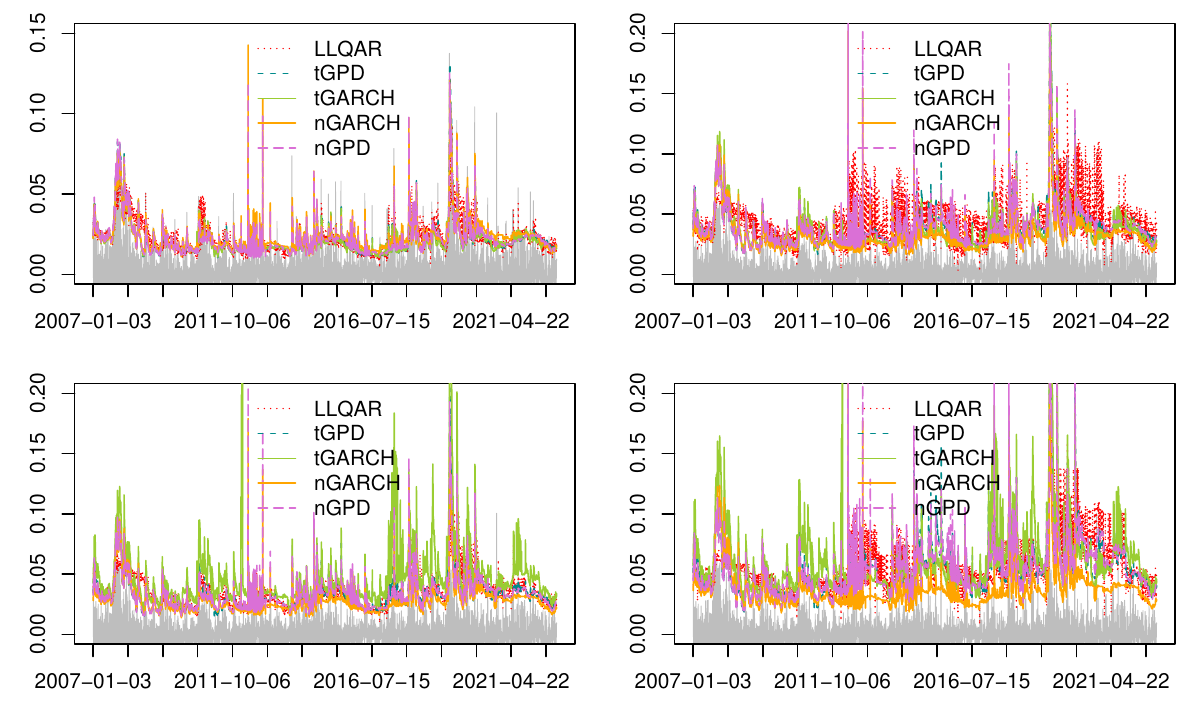}
       
      \caption{\textbf{Top}: Plot of $95\%$ (left) and $99\%$(right) VaR of different methods. \textbf{Bottom}: Plot of  $95\%$ (left) and $99\%$(right) ES of different methods for IBM Commom Stock data set.}\label{S-P-500-stock}
    \end{figure}

    \begin{figure}[H]
    \centering
     
       \includegraphics[width=4.5in]{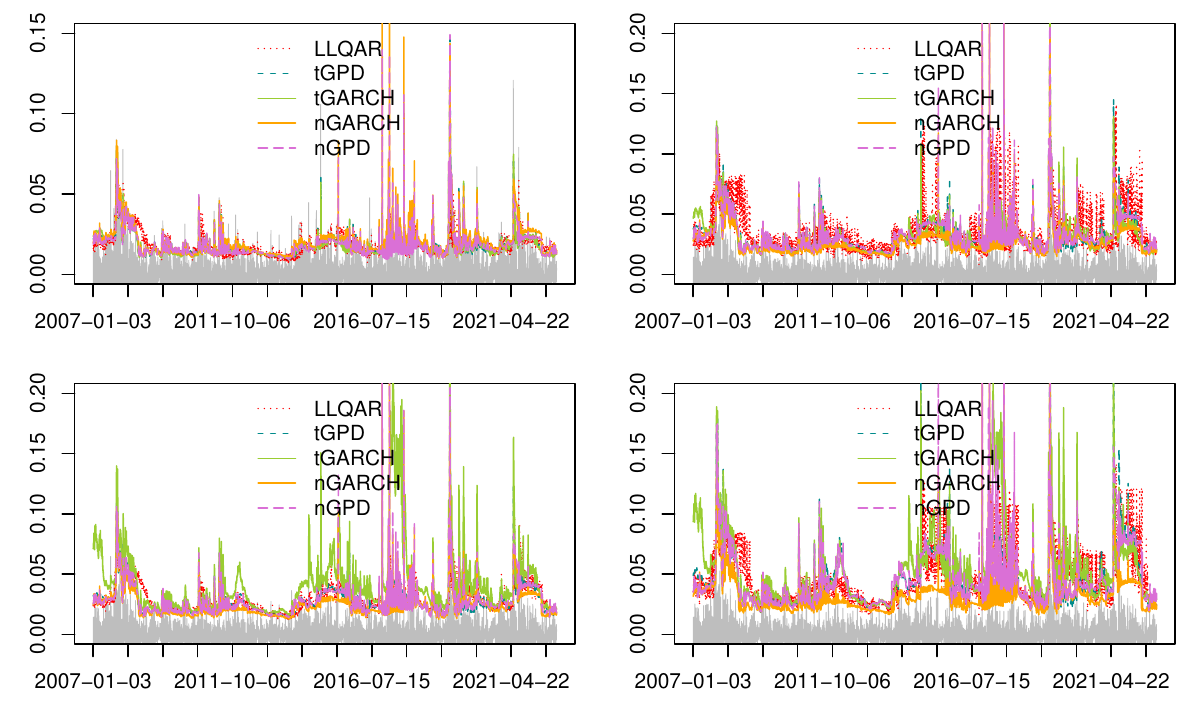}
       
      \caption{\textbf{Top}: Plot of $95\%$ (left) and $99\%$(right) VaR of different methods. \textbf{Bottom}: Plot of  $95\%$ (left) and $99\%$(right) ES of different methods for Walmart Stock data set.}\label{S-P-500-stock}
    \end{figure}

     The main motive of this research is to identify the model which has the best forecasting ability which is not possible by observing their forecasting plot. That's why we evaluated the performance of the methods by using formal backtesting procedures termed as Unconditional Coverage (UC) \cite{kupiec1995techniquesa} test and Conditional Coverage (CC) test  \cite{christoffersen1998evaluating}  
     for estimating VaR. Unfortunately, there are no significant research studies discharged to backtest ES.  This is also a possible door for further exploration. We conduct backtesting on ES proposed in \cite{mcneil2000estimation}. Following this we define the residual as,
     \begin{align}
         r_t&=\frac{L_t \mathbb{I}_{L_t>\widehat{VaR}_{\tau}}-\widehat{ES}}{\widehat{\sigma_t}}
     \end{align}
     where, $\widehat{\sigma_t}$ is the conditional SD of the model,  $\widehat{ES_{\tau}}$ is an estimate of $ES_{\tau}$, $\widehat{VaR_{\tau}}$ is an estimate of $VaR_\tau$  and $L_t$ is the log-losses. For a correctly specified model, the residuals $r_t$ should be IID and have zero mean. The testing procedure is done through a bootstrap technique. We use \textcolor{blue}{R} function ESTest from \textcolor{blue}{rugarch} package where they use one-sided t-test as well as bootstrap under the null hypothesis that the residuals $r_t$ is IID and has zero mean. For illustrating purposes later we specified this test as ES Bootstrap.
     Again for a better comparison, we introduced an additional measure inspired by the concept outlined in \cite{embrechts2005strategic} to backtest Expected Shortfall defined as:
    \begin{align}
        V_1&=\frac{\sum\limits_{t=1}^{T}\left(\widehat{ES}_{\tau} \mathbb{I}_{L_t>\widehat{VaR}_{\tau}}-\widehat{VaR}_{\tau}\mathbb{I}_{L_t>\widehat{VaR}_{\tau}}\right)}{\sum\limits_{t=1}^{T}1_{L_t>\widehat{VaR}_{\tau}}},\\
        V_2&=\frac{\sum\limits_{t=1}^{T}\left(L_t 1_{L_t>\widehat{VaR}_{\tau}}-\widehat{VaR}_\tau 1_{L_t>\widehat{VaR}_{\tau}}\right)}{\sum\limits_{t=1}^{T} 1_{L_t>\widehat{VaR}_{\tau}}},
    \end{align}
   where $V_2$ is the actual disparity between $L_t$ and $\widehat{VaR}_{\tau}$ when violation prevail and $V_1$ is the estimated disparity 
 and for a better estimate $V=V_1-V_2$ should be close to 0.
 
 Tables \ref{table-3} and \ref{table-4} report the $p$-value of three backtesting results, the proportion of violation and the value of $V$ for two $\tau$ levels for four stocks stated above. We will reject the null hypothesis if the $p$-value is less than 0.05 for both quantiles. For each method, the favorable results are in bold.

  \begin{table}
 \scalebox{1}{
    \begin{tabular}{c|c|c|c|c}\hline
 Data &Methods & UC (PV)  &  CC  & $V$[ES Bootstrap]\\\hline

 
  
 & tGARCH & \textbf{0.0521} [0.0569] & 0.0495 &0.00729[\textbf{1.0000}]\\

 & nGARCH & \textbf{0.097} [0.0558] & \textbf{0.0987} &-0.0028[9.62e-08]\\

 & GPD-nGARCH & \textbf{0.7123} [0.0512] & \textbf{0.3260 }&-0.00018[\textbf{0.3744}]\\
 
S \& P 500 & GPD-tGARCH & \textbf{0.9444}  [0.0497] & \textbf{0.2477} &-0.00028\textbf{[0.3149]}\\

& LLQAR & \textbf{0.3349} [0.0466]& \textbf{0.6206} & \textbf{-0.00013}[\textbf{0.4461}]\\

& SAV &7.37e-05[0.064] & 2.27e-04\\
& AS &9.e-06[0.07] & 1.14e-05\\
& IN-GARCH &1.53e-05[0.066] & 1.06e-05\\
& ADAPTIVE  & 6.37e-04[0.062] & 9.71e-09\\\hline


& tGARCH & \textbf{0.9387} [0.0502] & 0.00461&0.0125[\textbf{1.0000}]\\

 & nGARCH & \textbf{0.0618} [0.0436] & 0.00281 &-0.0058[4.28e-06]\\

 & GPD-nGARCH & 9.97e-01 [0.05001] & 0.00162 &-0.00193[\textbf{0.053}]\\
 
Apple Stock & GPD-tGARCH & \textbf{0.381}  [0.0530] & 9.56e-04&-0.0011\textbf{[0.160]}\\

& LLQAR & \textbf{0.767} [0.0510]& \textbf{0.881} & 0\textbf{.00037[0.6103}]\\
& SAV & \textbf{0.1927}[0.0546] & 2.14497e-05 &\\
& AS &2.271e-05 [0.0655] &6.56e-08 &\\
& IN-GARCH &\textbf{0.1838}[0.055] & 1.65e-04\\
& ADAPTIVE  & \textbf{0.192}7[0.055] & 4.28e-04\\\hline

& tGARCH & \textbf{0.9387} [0.0502] & 0.0485&0.0125[\textbf{1.0000}]\\

 & nGARCH & 0.0129 [0.0415] & 0.0037&-0.00753[4.94e-10
]\\

 & GPD-nGARCH & \textbf{0.7673} [0.0510] & 0.00107 &-0.0013[\textbf{0.1134}]\\
 
IBM Common Stock & GPD-tGARCH & \textbf{0.6073}  [0.0517] & 0.0036&\textbf{-0.00057[0.2962]}\\

& LLQAR & \textbf{0.6073} [0.0518]& \textbf{0.3387} & 0.000897[\textbf{0.7789}]\\

& SAV & 2.93e-04[0.063] & 1.69e-07 &\\
& AS &4.47e-10 [0.073] &2.78e-15 &\\
& IN-GARCH &0.0048 [0.06] & 5.07e-06\\
& ADAPTIVE  & 0.0021[0.061] & 9.09e-08\\\hline

& tGARCH & \textbf{0.4220} [0.0528] & \textbf{0.2271}&0.00745[\textbf{1.0000}]\\

 & nGARCH & 0.0243 [.04235] & 0.00175 &-0.0065[6.48e-07]\\

 & GPD-nGARCH & \textbf{0.659} [0.0515] & \textbf{0.125} &-0.0022[\textbf{0.1389}]\\
 
Walmart Stock & GPD-tGARCH & \textbf{0.8284}  [.0492] & \textbf{0.1227}&-0.0025[0.4962]\\

& LLQAR & \textbf{0.659} [0.0515]& \textbf{0.485} & \textbf{-0.0012}[\textbf{0.7808}]\\
& SAV & 0.0712 [.056]& 0.0102 &\\
& AS &  7.63e-09[0.07] & 1.764888e-11 \\
& IN-GARCH &\textbf{0.07924} [0.056] & 0.0053\\
& ADAPTIVE  & \textbf{0.061}[0.057] & 0.0026\\\hline

\end{tabular}}  
\caption{ For $\tau=0.05$, the p-value of the UC and CC test for  VaR methodologies are given in the first two columns with the proportion of violation (PV) under the UC column within the third bracket are displayed and the value of $V$ and p-value of ES Bootstrap test defined earlier evaluating ES by various methods is given in the last column for S \& P 500 Index, Apple Stock, IBM Common Stock and Walmart stock. For each method, the favorable result is shown in bold text.}\label{table-3}
\end{table}


\begin{table}
 \scalebox{1}{
    \begin{tabular}{c|c|c|c|c}\hline
 Data &Methods & UC (PV)  &  CC   & $V$[ES Bootstrap]\\\hline

 
  
 & tGARCH & 8.845e-05 [0.0168] & 9.605e-06 &0.0055[\textbf{0.999}]\\

 & nGARCH & 8.726e-14 [0.0239] & 2.675e-14 
 &-0.0035[-0.0045]\\

 & GPD-nGARCH & 0.0167[0.0140] & 2.445e-04 &-0.0019[0.032688]\\
 
S \& P 500 & GPD-tGARCH & \textbf{0.052  [.0132]} & 3.57e-05 &\textbf{-0.00158}[0.1074]\\

& LLQAR & 1.535e-04  [0.0161]& 6.623e-05 
& 0.026[\textbf{0.999}]\\

& SAV &0[0.03] & 0\\
& AS & 0[0.039] & 0\\
& IN-GARCH & 0[0.03] & 0\\
& Adaptive  & 0[0.3] & 0\\\hline


& tGARCH & \textbf{0.096} [0.01275] & \textbf{0.1311} &0.0092[\textbf{0.998}]\\

 & nGARCH & 4.432e-04 [0.0161] & 7.467e-04 &-0.0099[5.84e-05]\\

 & GPD-nGARCH & 0.0353 [0.0135] & \textbf{0.0528} &-0.0064[0.0087]\\
 
Apple Stock & GPD-tGARCH & 0.0245  [0.0137] & 0.0375 &\textbf{-0.0055}[0.0176]\\

& LLQAR & 0.0167 [0.0140]& 0.0261& 0.033[\textbf{0.999}]\\
& SAV &1.799e-08[0.02] & 5.995e-10\\
& AS & 0[0.037] & 0\\
& IN-GARCH & 1.45e-13 [0.024] & 3.44e-13\\
& Adaptive  & 0.012[0.014] & 0.04\\\hline

& tGARCH & 0.0501 [0.01326] & \textbf{0.0626} &0.0064[\textbf{0.975}]\\

 & nGARCH & 8.439e-06  [0.0178] & 4.0439e-05 
 &-0.012[9.868e-11]\\

 & GPD-nGARCH & 0.0167[0.01403] & 0.0282&\textbf{-0.0022[0.1757}]\\
 
IBM common Stock & GPD-tGARCH & \textbf{0.2240  [0.0119]} & \textbf{0.1525} &-0.0049[0.0231]\\

& LLQAR & 0.02452 [ 0.0137]& 0.0375& 0.042[\textbf{0.998}]\\
& SAV &3.995e-09[0.021] &3.54e-11\\
& AS & 0[0.043] & 0\\
& IN-GARCH & 1.38e-09 [0.021] & 1.56e-11\\
& Adaptive  & 1.54e-04[0.017] & 2.28e-06 \\\hline


& tGARCH & 0.0167 [0.0140] & 0.00164 &\textbf{0.0024}[\textbf{0.8236}]\\

 & nGARCH & 6.464e-07  [0.0140] & 4.933e-08 &-0.0089[6.133e-06]\\

 & GPD-nGARCH & \textbf{0.07} [0.0130] & \textbf{0.0783 }&-0.0051[0.0318]\\
 
Walmart Stock & GPD-tGARCH & \textbf{0.224  [0.0119]} & \textbf{0.1525} &-0.0066[0.0171]\\

& LLQAR & 0.0245 [0.0137]& \textbf{0.07657}& 0.028[\textbf{1}]\\
& SAV & 4.35e-13[.024]&2.22e-16& \\
& AS & 0 [.04]& 0 & \\
& IN-GARCH & 1.95e-12 [0.023] & 1.83e-14 \\
& Adaptive  & 1.53e-05[0.018] & 7.53e-08 \\\hline
\end{tabular}}  
\caption{ For $\tau=0.01$, the p-value of the UC and CC test for  VaR methodologies are given in the first two columns with the proportion of violation (PV) under the UC column within the third bracket are displayed and the value of $V$ and p-value of ES Bootstrap test defined earlier evaluating ES by various methods is given in the last column for S \& P 500 Index, Apple Stock, IBM Common Stock and Walmart stock. For each method, the favorable result is shown in bold text.}\label{table-4}
\end{table}

In Table \ref{table-3}, the results for the 95\% quantile reveal that the proportion of violations is closest to $\tau$ for all stocks and indexes across all methods, except for GPD-nGARCH, which exhibits a high tendency for underestimation. UC (Unconditional Coverage Test) is passed by all methods, excluding CAViaR models for S\&P 500 data, GPD-nAGRCH, and AS methods (see Appendix C) for Apple Stock, GPD-nGARCH and CAViaR models for IBM Common Stock, and all methods except nGARCH, SAV, and AS for Walmart data. The results also indicate clustering of violations, as the CC (Correct Exceedances and Independence test) is rejected by almost all methods for all datasets, except LLQAR, nGARCH, GPD-nGARCH, and GPD-tGARCH, which pass for S\&P 500. Only LLQAR passes the CC test for Apple and IBM Common Stocks. For Walmart data, all methods except nGARCH and CAViaR models pass the tests for 95\% quantile estimates, suggesting that GARCH models tend to perform less accurately than LLQAR at lower quantiles.
Table \ref{table-4}, presenting the 99\% quantile estimates, shows that all methods are rejected in both UC and CC tests, with GPD-tGARCH and tGARCH slightly performing better in backtesting. For S\&P 500 data, only GPD-tGARCH passed the UC test, and no other method passed the CC test. For Apple Stock, the only method passing both tests is tGARCH, indicating its superior performance. Similarly, for IBM Common stock, GPD-tGARCH is the only method passing both tests. In the case of Walmart Stock, the UC test is passed by GPD-nGARCH and GPD-tGARCH, and for CC tests, the passing methods are GPD-nGARCH, GPD-tGARCH, and LLQAR. It is concluded that the LLQAR method performs better than GARCH methods for 95\% VaR, while GPD-tGARCH provides more accurate and reliable estimates for 99\% VaR in most datasets.
The best-performing VaR model does not necessarily lead to the best ES estimation. The last column in both tables displays the value of V and the p-value of ES Bootstrap defined earlier. The results depict that all methods pass the ES Bootstrap test as the p-value is greater than 0.05 except nGARCH for the 95\% quantile. Again, for $99\%$ estimates, only LLQAR and tGARCH pass the ES Bootstrap test except IBM Common Stock, where GPD-nAGRCH also passes this test. It gives clear evidence that the exceedance residuals extracted from all methods are IID and have zero means except for the failure methods. The smallest value of $V$ is found for LLQAR for S\&P 500 and Apple and Walmart Stock for the 95\% quantile, but for IBM Common Stock it is found for GPD-tGARCH. Again, it is observed that for 99\% of estimates, LLQAR has the smallest V value except for IBM common stock, where GPD-tGARCH has the smallest value. It is evident that tGARCH, GPD-tGARCH, and LLQAR act similarly well based on both evaluation methods for all stocks. It gives a supportive indication that non-parametric LLQAR is competitive and sometimes performs slightly better than GARCH family models for ES estimates.
\\

It is evident that for both quantiles, tGARCH and GPD-tGARCH perform the best for all stocks, while LLQAR emerges as the second-best-performing model for both VaR and ES estimation. This conclusion aligns with our findings from the simulation study, particularly in non-stationary cases.

\section{Discussion}

This study centers around the estimation of Value at Risk (VaR) and Expected Shortfall (ES). In this paper, we introduced a novel non-parametric approach, the Local Linear Quantile Autoregression (LLQAR) model, for VaR and ES estimation. To facilitate this, we devised a new weight function and outlined a procedure for bandwidth selection. Additionally, we discussed parametric, semi-parametric, and non-parametric approaches, presenting VaR and ES formulas for each of these methods.

The justification of the performance of the proposed model and its comparison with other benchmark models is established through a comprehensive simulation study and application to real-world data. The evaluation metrics, such as Root Mean Square Error (RMSE) for the simulation study and traditional backtesting procedures for both the simulation study and real-world data, are employed to compare the performances. While our proposed method did not emerge as the overall winner, it demonstrated competitiveness and showcased the best performance in certain cases, particularly for non-stationary data.


Additionally, it was observed that the proposed method consistently provided estimates very close to those of the best-performing methods in most cases. It's important to note that this study aimed to compare the proposed non-parametric model with benchmark models, and the intention is not to recommend the proposed method as the best-performing model. Instead, it serves as a justification for utilizing the new non-parametric method, LLQAR, for risk estimation, with room for improvement by interested researchers who may consider various insightful characteristics.

An identified limitation of the LLQAR method is its tendency to decay slowly after consecutive large shocks, leading to overestimation when compared to benchmark GARCH models. This phenomenon is a well-known volatility clustering effect in financial time series, and addressing this situation may require a deeper analysis of spatial and temporal dependencies.


A critical limitation of the LLQAR method lies in the challenge of selecting the optimal smoothing parameter, which greatly influences its implementation and the quality of estimates. This presents a potential avenue for future research to explore ways to determine the best-performing smoothing parameter that can enhance the method's efficacy.

Another potential issue in LLQAR, although not observed in the current data analysis, is that of quantile crossing during the estimation of various quantiles in both VaR and ES. Future studies could investigate strategies to avoid quantile crossing and enhance the methodology. One possible direction is to explore the combination of LLQAR with methods designed to prevent quantile crossing, such as the approach presented in \cite{bondell2010noncrossing}.

Additionally, the current study considered only past losses as predictors. Future research could explore alternative approaches, such as incorporating squared losses and the absolute value of losses as predictors, leading to multiple LLQAR models. This exploration may contribute to improving the performance of VaR and ES estimation.

\section*{Disclosure statement}

To the best of our knowledge there is no research conflict with our work. 

\bibliography{Var}
\clearpage
\appendix
\section{\\A.1 Estimating Value at Risk (VaR) and Expected Shortfall (ES) in the third scenario of the simulation study.}
The Value-at-Risk (VaR) and Expected Shortfall (ES) estimations for scenario 3 in the simulation study are illustrated in the following plots. These graphs provide insights into the risk assessment outcomes under the specified conditions of scenario 3.

\begin{figure}[H]
    \centering
    
   \includegraphics[width=3.4in]{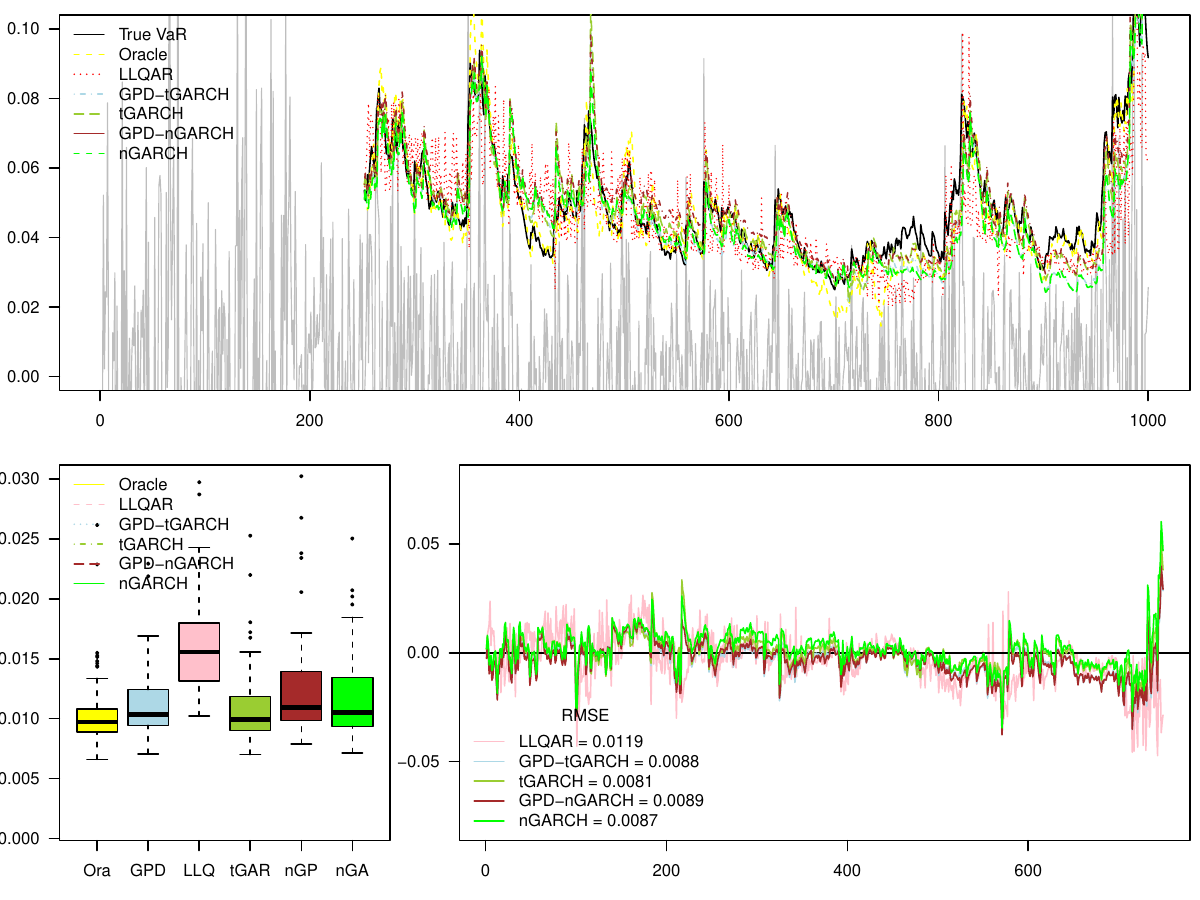}
     \caption{\textbf {Top:} One step ahead 95\% VaR predictions from a simulated data. \textbf{Bottom left:} Box plots of RMSE.  \textbf{Bottom right:} The VaR prediction error series for the simulated series above (Stationary dataset).}\label{one-percent}
\end{figure}

\begin{figure}[H]
       \centering
       \includegraphics[width=3.4in]{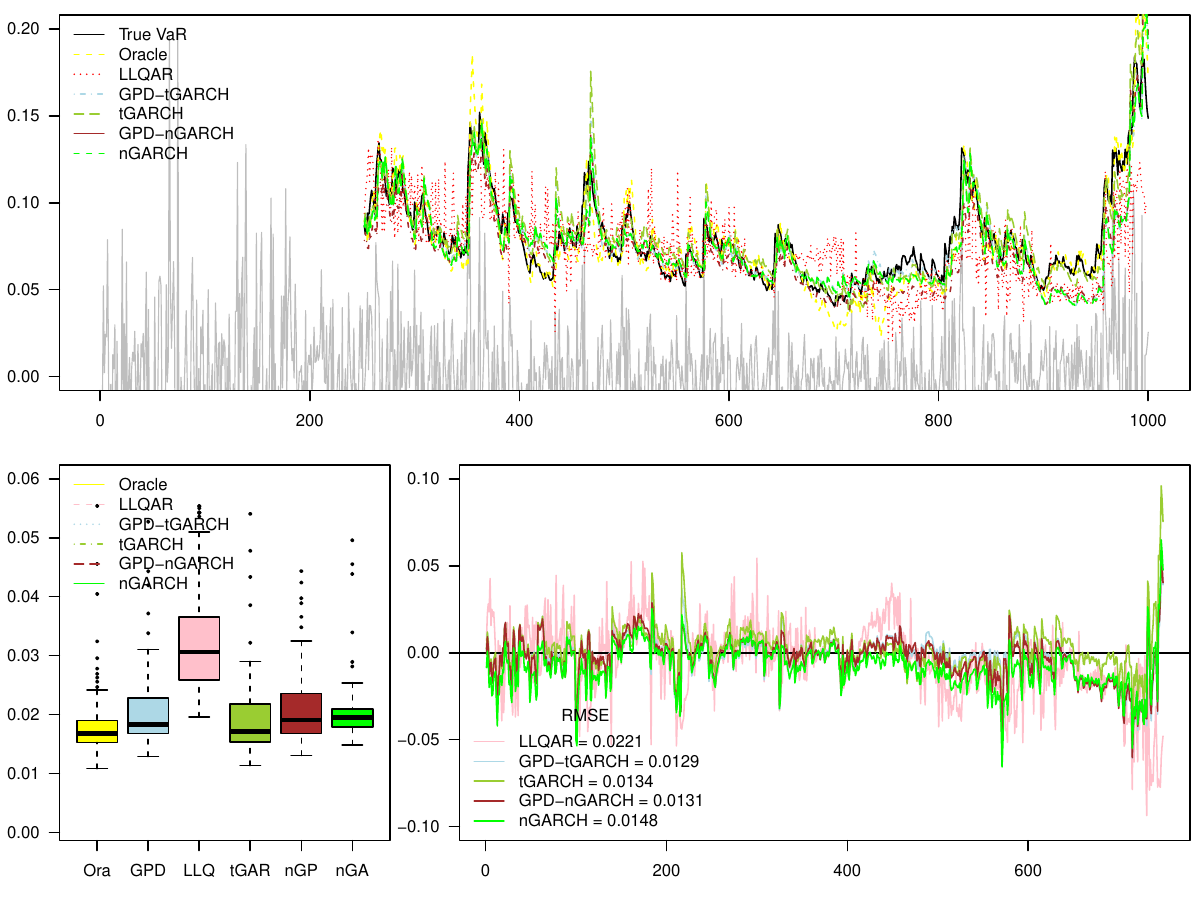}
       \caption{\tiny\textbf {Top:} One step ahead 99\% VaR forecast. \textbf{Bottom:} Box plot of RMSE (left) and corresponding error (right) plot of simulated data with  RMSE (Stationary dataset).}\label{two-percent}
\end{figure}

\begin{figure}[H]
    \centering
    \includegraphics[width=3.4in]{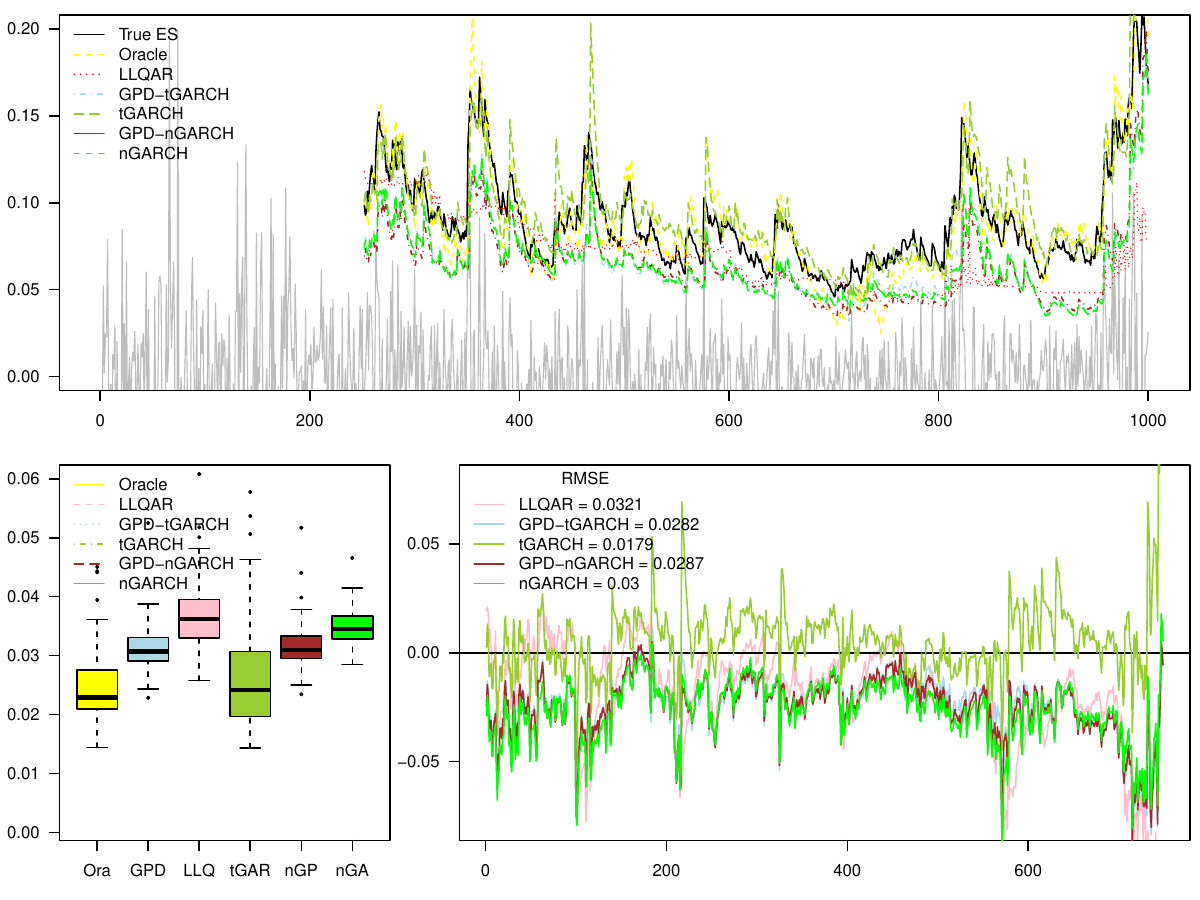}
     \caption{\tiny\textbf {Top:} One step ahead 95\% ES forecast. \textbf{Bottom:} Box plot of RMSE (left) and corresponding error (right) plot of simulated data with  RMSE (Stationary dataset).}
\end{figure}

\begin{figure}[H]
       \centering
       \includegraphics[width=3.4in]{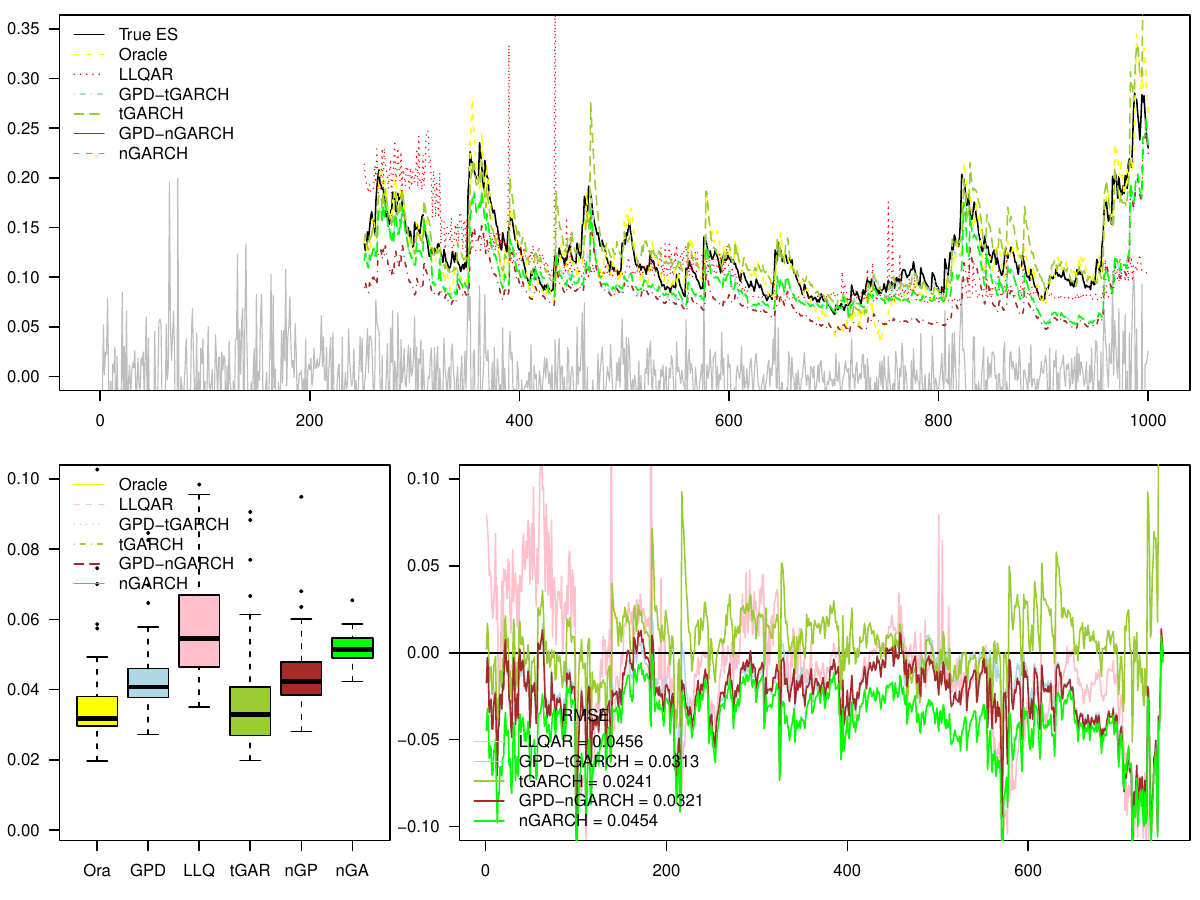}
       \caption{\tiny\textbf {Top:} One step ahead 99\% ES forecast. \textbf{Bottom:} Box plot of RMSE (left) and corresponding error (right) plot of simulated data with  RMSE (Stationary dataset).}
\end{figure}

\begin{figure}[H]
    \centering
    \includegraphics[width=3.4in]{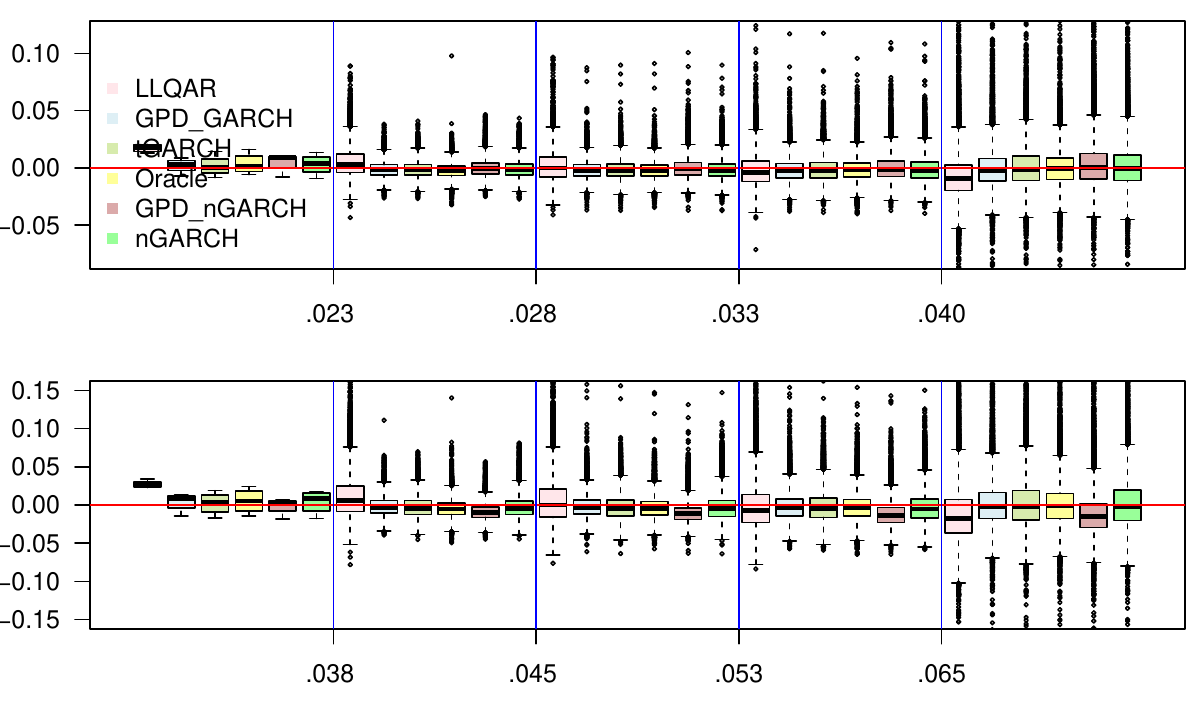}
     \caption{\tiny\textbf {Top:} Box plot of error of 95\% VaR forecast. \textbf{Bottom:} Box plot of error of 99\% VaR forecast (Stationary dataset).}
\end{figure}

\begin{figure}[H]
    \centering
    \includegraphics[width=3.4in]{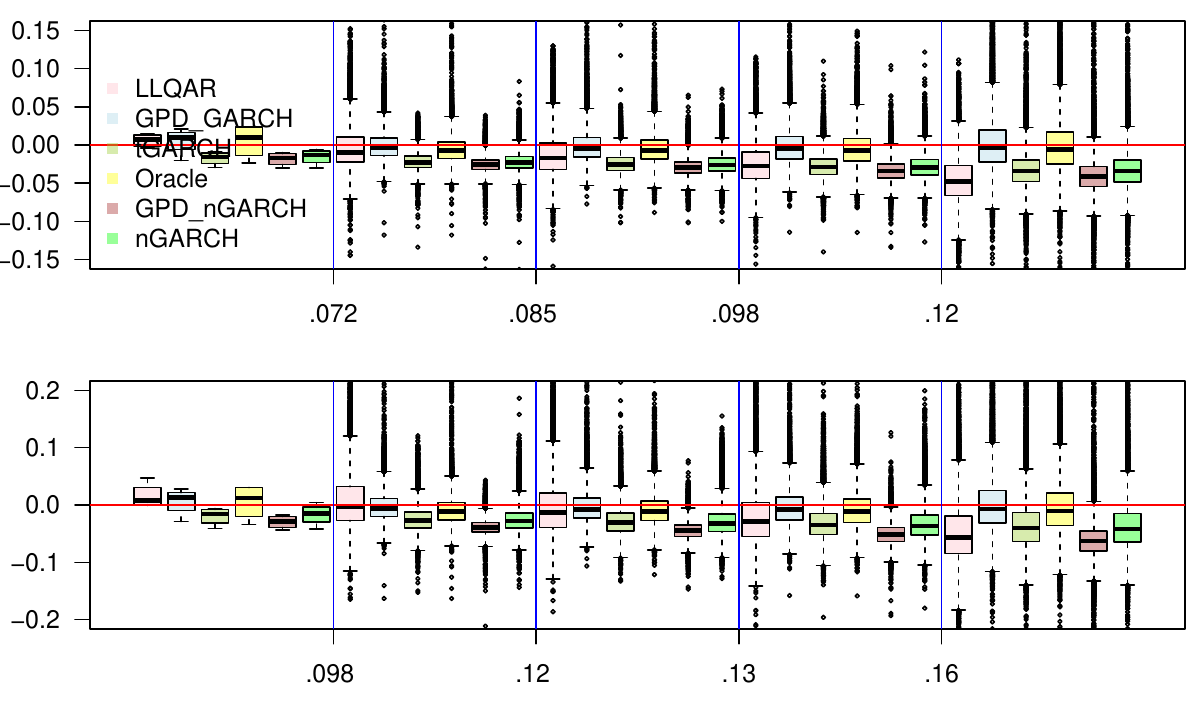}
     \caption{\tiny\textbf {Top:} Box plot of error of 95\% ES forecast. \textbf{Bottom:} Box plot of error of 99\% ES forecast (Stationary dataset).}
\end{figure}

\section{\\B.1 VaR estimation plot of CAViaR models.} The Conditional Autoregressive Value at Risk (CAViaR) model, as introduced by Engle and Manganelli \cite{engle2004caviar}, estimates quantiles over time through an autoregressive process. The CAViaR model takes four different model forms, which are outlined below. :\\
 Symmetric Absolute Value (SAV) model : $Q_t(\beta)=\beta_1+\beta_2Q_{t-1}(\beta)+\beta_3|y_{t-1}|$\\

 Asymmetric (AS) model : $Q_t(\beta)=\beta_1+\beta_2Q_{t-1}(\beta)+\beta_3(y_{t-1})^{+}-\beta_4 (y_{t-1})^{-}$\\

 In-direct-GARCH model : $Q_t(\beta)=\beta_1+\beta_2Q_{t-1}^2(\beta)+\beta_3(y_{t-1})^2$\\

 Adaptive model : $Q_t(\beta)=Q_{t-1}(\beta_1)+\beta_1\{[1+exp(G[y_{y-1}-Q_{t-1}(\beta_1)])]^{-1}-\theta\}$,\\
\\
where, G is some positive finite number. For finite G, this
model is a smoothed version of a step function.
The parameters with regression quantiles.\\
 The performances of CAViaR models were compared with the parametric models and non-parametric LLQAR, focusing only on real-world applications. It was predominantly observed that the performances of these models were inferior to LLQAR and parametric models. The convergence rate for simulated datasets with CAViaR models was very slow, which is why those results were omitted. The graphical presentation of the performances of different methods is shown below:\\
 
\begin{figure}[H]
    \centering
     
       \includegraphics[width=4.5in]{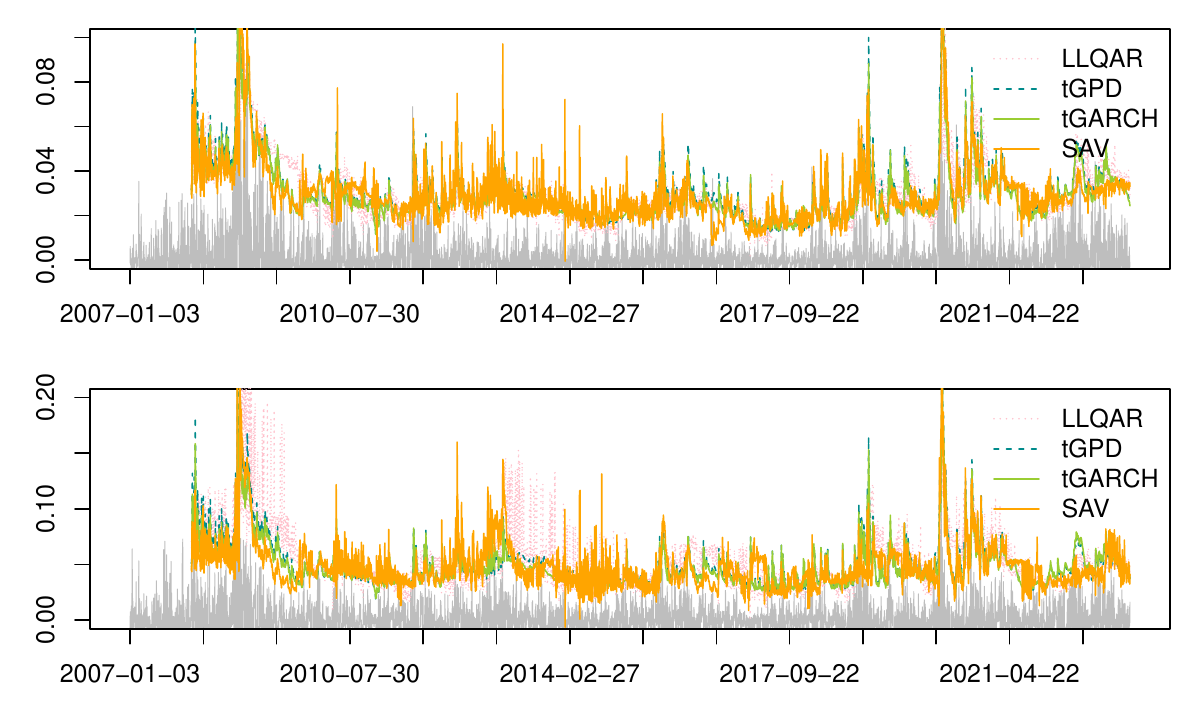}
       
      \caption{\textbf{Top}: Plot of $95\%$ (top) and $99\%$(bottom) VaR of different methods of APPL with SAV method.}
    \end{figure}

    \begin{figure}[H]
    \centering
     
       \includegraphics[width=4.5in]{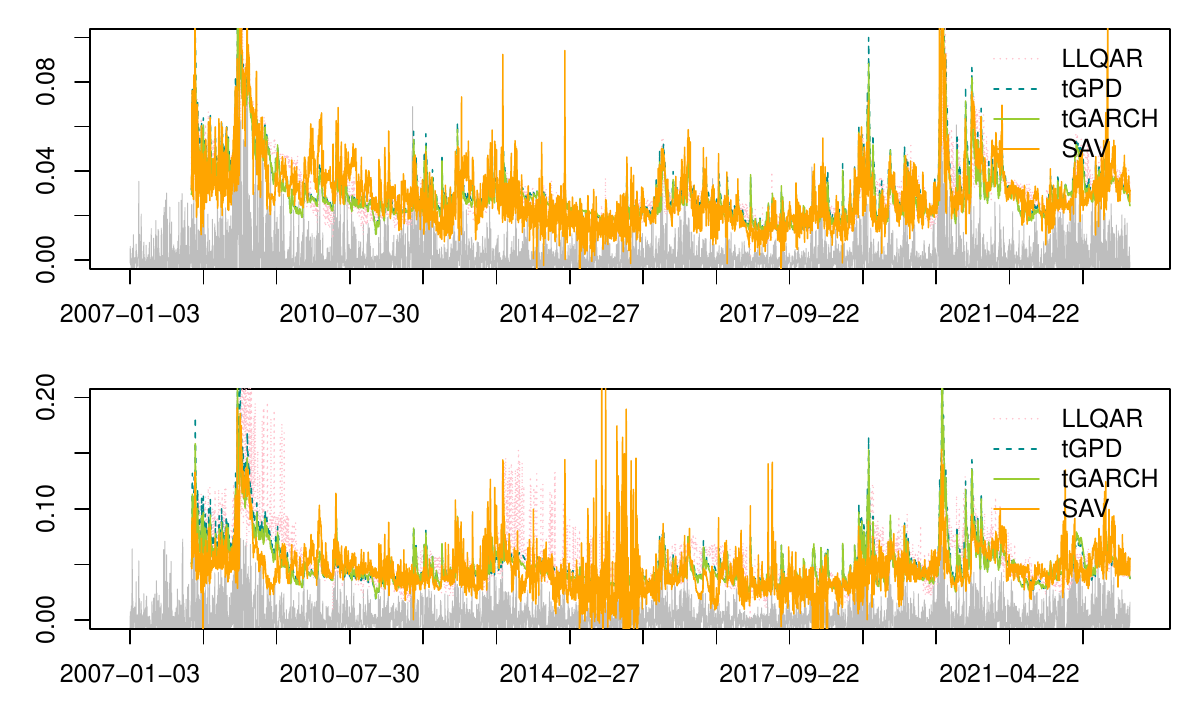}
       
      \caption{\textbf{Top}: Plot of $95\%$ (top) and $99\%$(bottom) VaR of different methods of APPL with AS method.}
    \end{figure}
    
    \begin{figure}[H]
    \centering
     
       \includegraphics[width=4.5in]{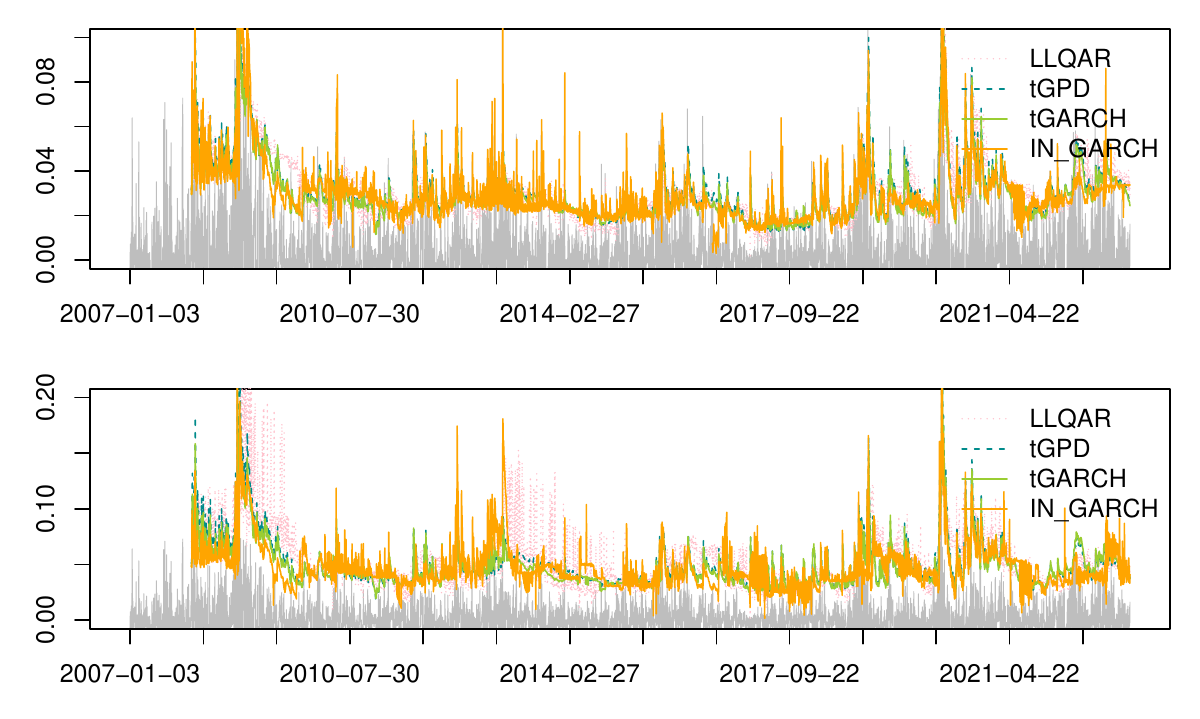}
       
      \caption{\textbf{Top}: Plot of $95\%$ (top) and $99\%$(bottom) VaR of different methods of APPL with In-direct-GARCH method.}
    \end{figure}

    \begin{figure}[H]
    \centering
     
       \includegraphics[width=4.5in]{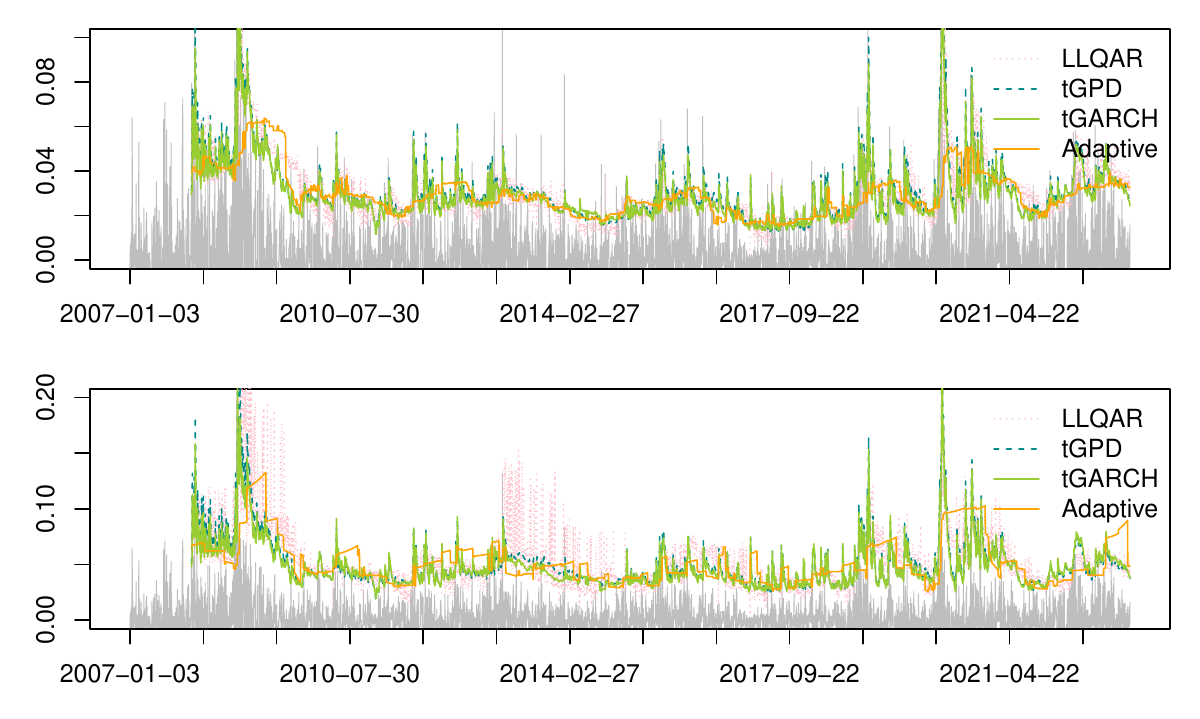}
       
      \caption{\textbf{Top}: Plot of $95\%$ (top) and $99\%$(bottom) VaR of different methods of APPL with Adaptive method.}
    \end{figure}
    
    
    \begin{figure}[H]
    \centering
     
       \includegraphics[width=4.5in]{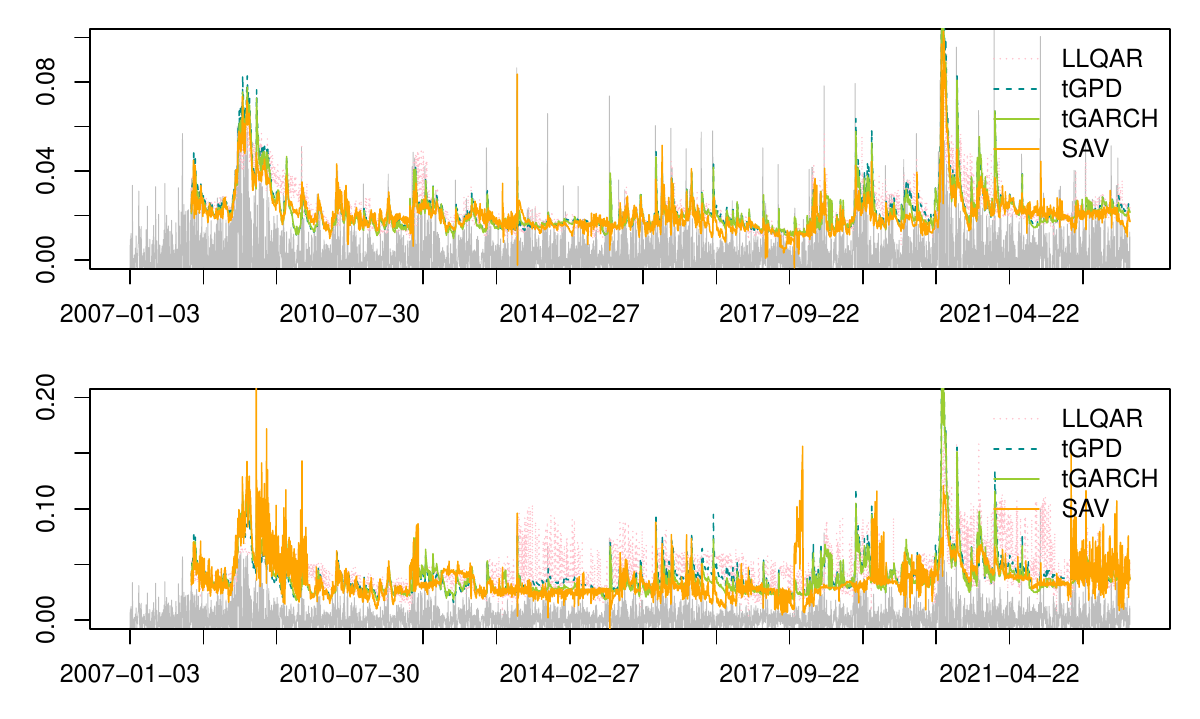}
       
      \caption{\textbf{Top}: Plot of $95\%$ (top) and $99\%$(bottom) VaR of different methods of IBM with SAV method.}
    \end{figure}

    \begin{figure}[H]
    \centering
     
       \includegraphics[width=4.5in]{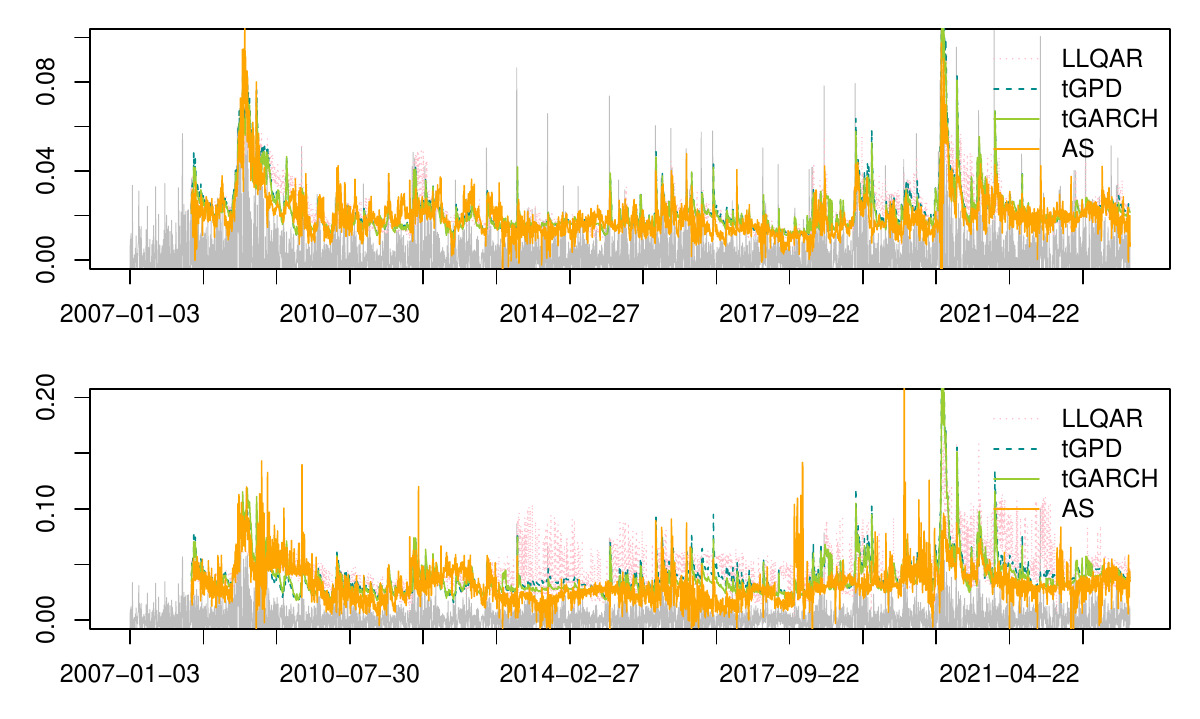}
       
      \caption{\textbf{Top}: Plot of $95\%$ (top) and $99\%$(bottom) VaR of different methods of IBM with AS method.}
    \end{figure}
    
    \begin{figure}[H]
    \centering
     
       \includegraphics[width=4.5in]{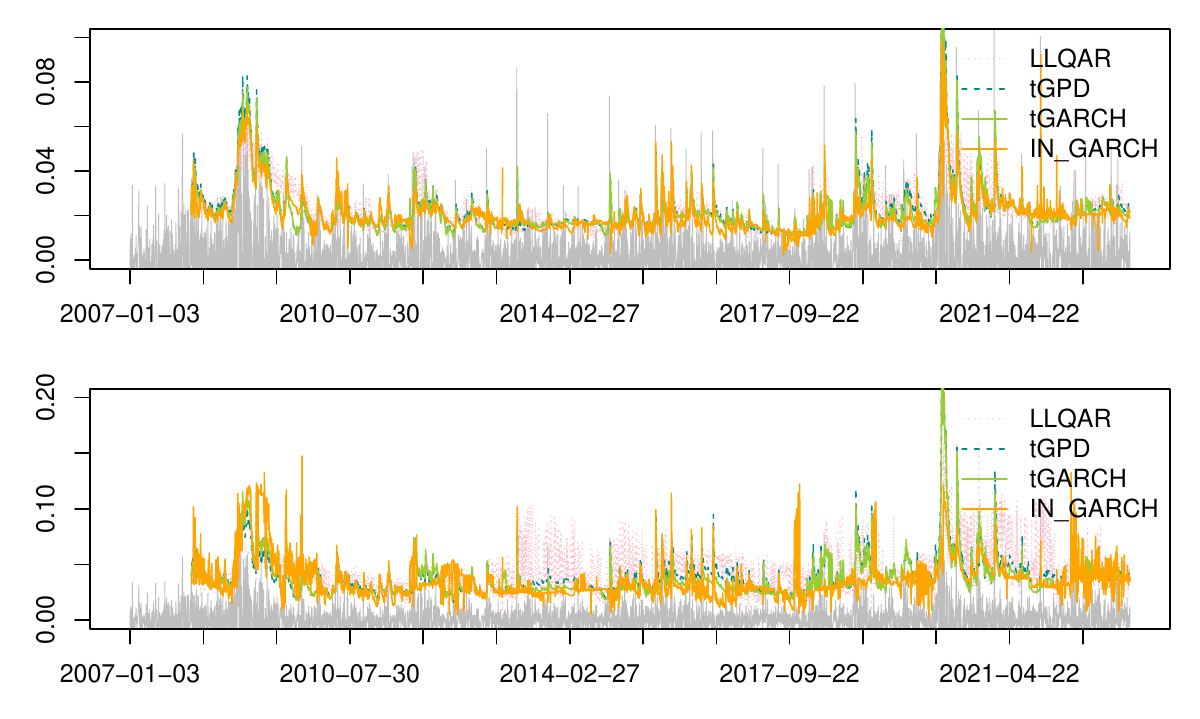}
       
      \caption{\textbf{Top}: Plot of $95\%$ (top) and $99\%$(bottom) VaR of different methods of IBM with In-direct-GARCH method.}
    \end{figure}

    \begin{figure}[H]
    \centering
     
       \includegraphics[width=4.5in]{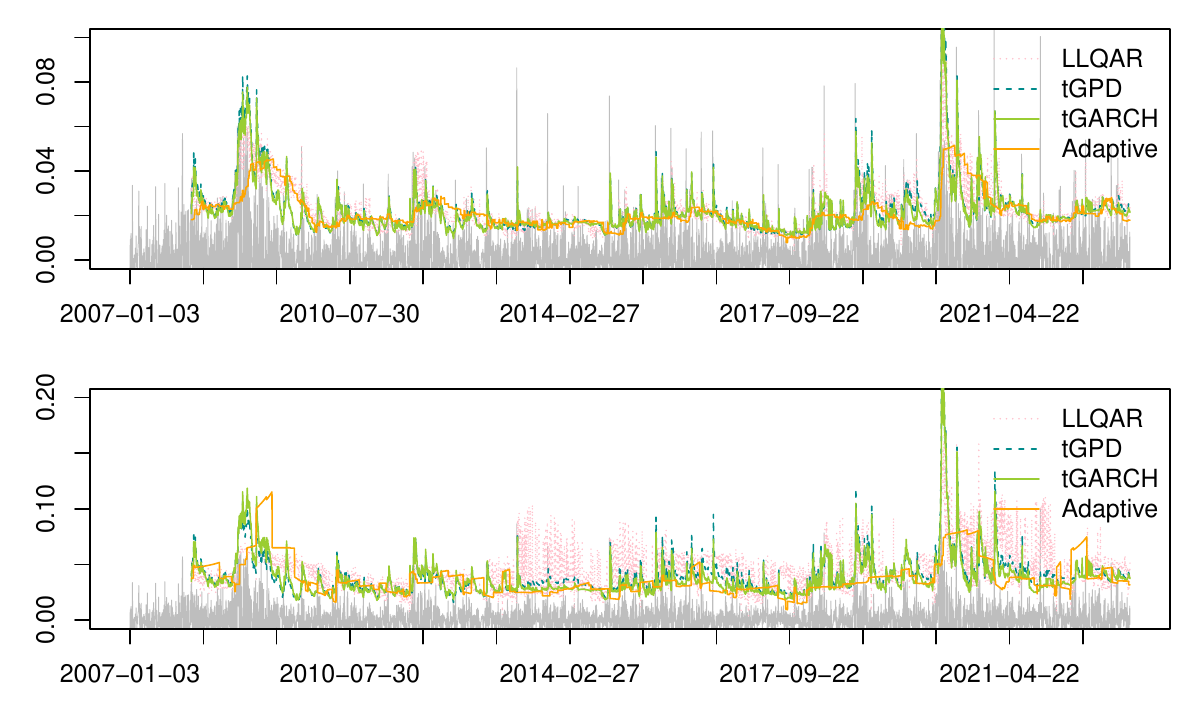}
       
      \caption{\textbf{Top}: Plot of $95\%$ (top) and $99\%$(bottom) VaR of different methods of IBM with Adaptive method.}
    \end{figure}
    

\begin{figure}[H]
    \centering
     
       \includegraphics[width=4.5in]{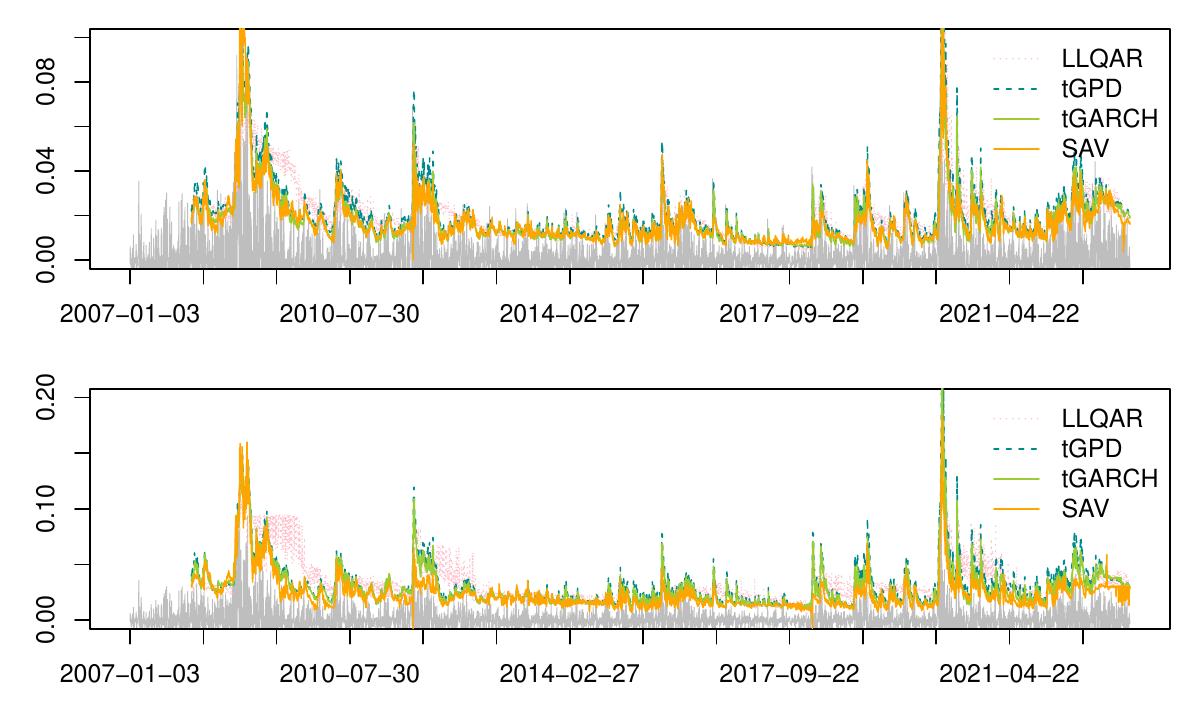}
       
      \caption{\textbf{Top}: Plot of $95\%$ (top) and $99\%$(bottom) VaR of different methods of SP with SAV method.}
    \end{figure}

    \begin{figure}[H]
    \centering
     
       \includegraphics[width=4.5in]{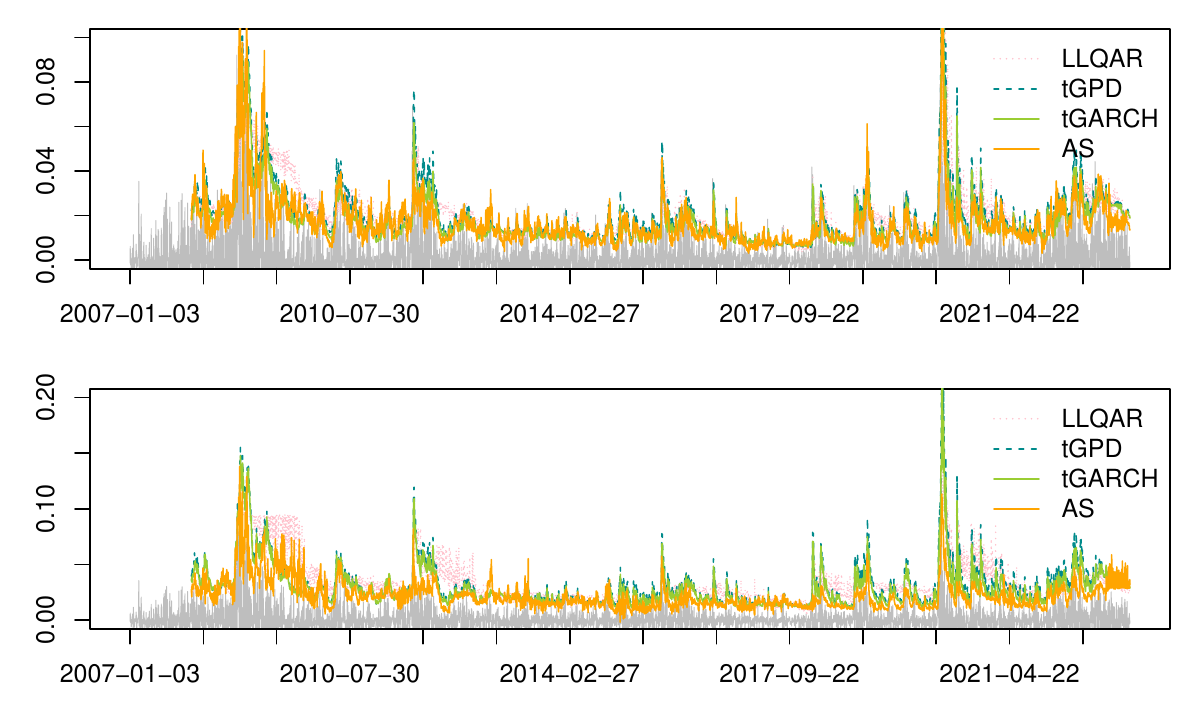}
       
      \caption{\textbf{Top}: Plot of $95\%$ (top) and $99\%$(bottom) VaR of different methods of SP with AS method.}
    \end{figure}
    
    \begin{figure}[H]
    \centering
     
       \includegraphics[width=4.5in]{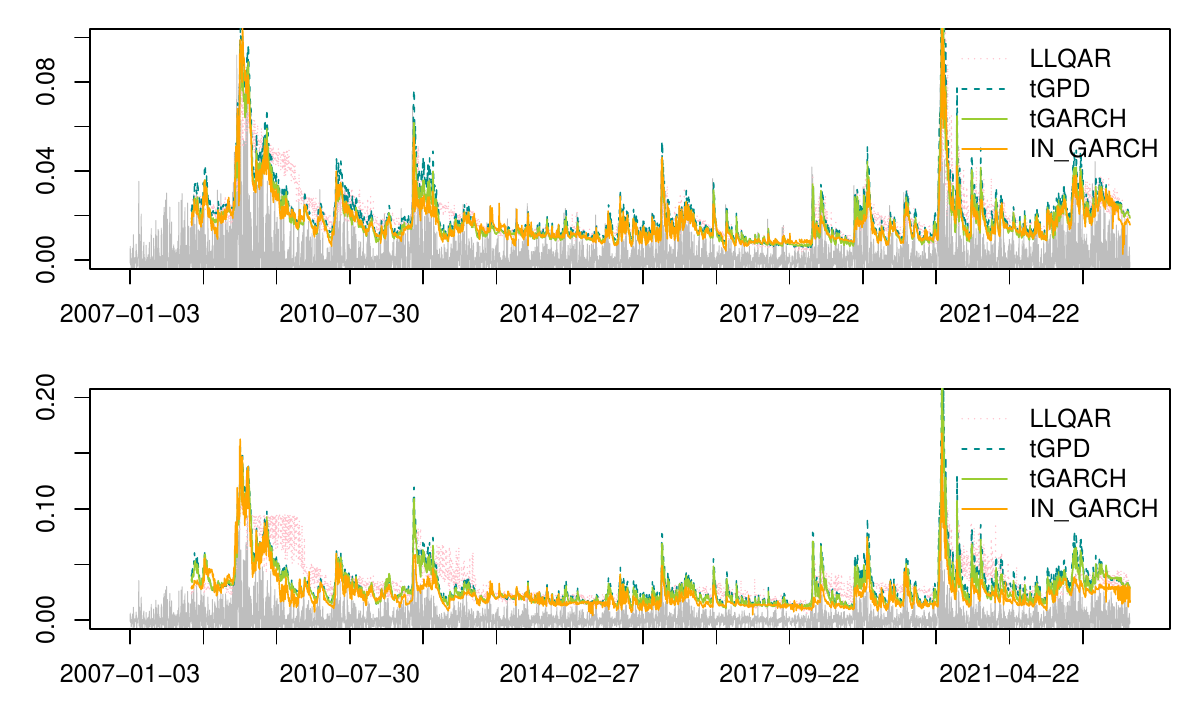}
       
      \caption{\textbf{Top}: Plot of $95\%$ (top) and $99\%$(bottom) VaR of different methods of SP with In-direct-GARCH method.}
    \end{figure}

    \begin{figure}[H]
    \centering
     
       \includegraphics[width=4.5in]{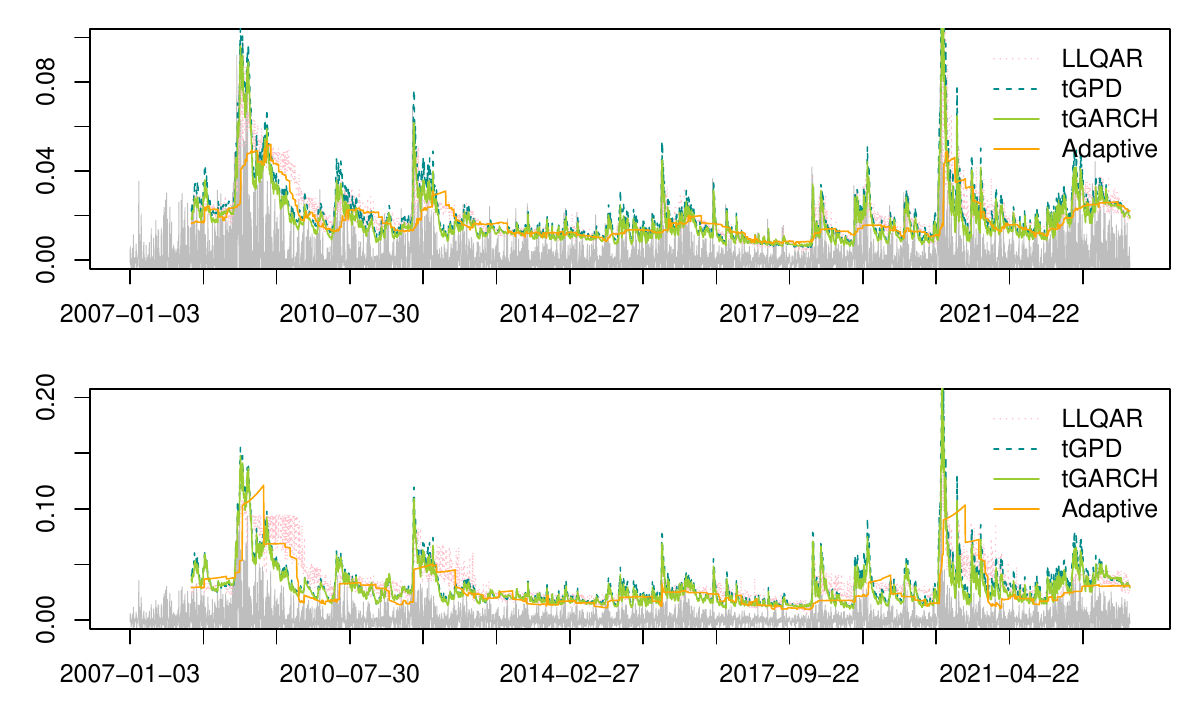}
       
      \caption{\textbf{Top}: Plot of $95\%$ (top) and $99\%$(bottom) VaR of different methods of SP with Adaptive method.}
    \end{figure}
    

\begin{figure}[H]
    \centering
     
       \includegraphics[width=4.5in]{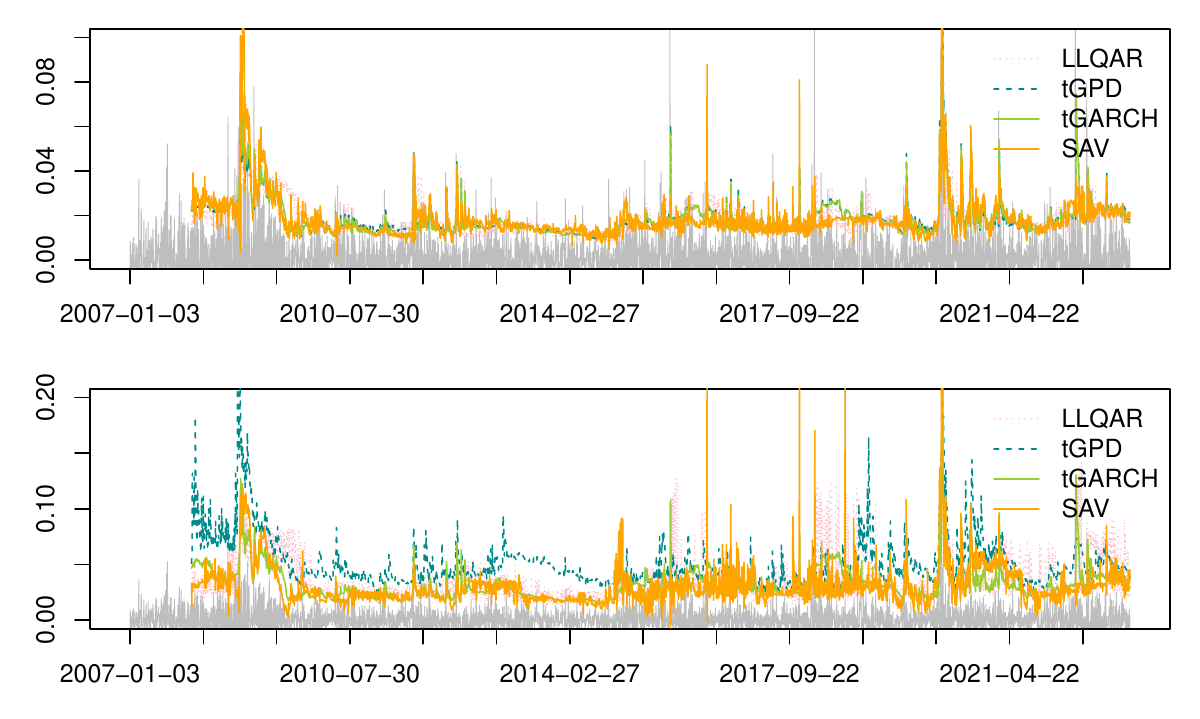}
       
      \caption{\textbf{Top}: Plot of $95\%$ (top) and $99\%$(bottom) VaR of different methods of WMT with SAV method.}
    \end{figure}

    \begin{figure}[H]
    \centering
     
       \includegraphics[width=4.5in]{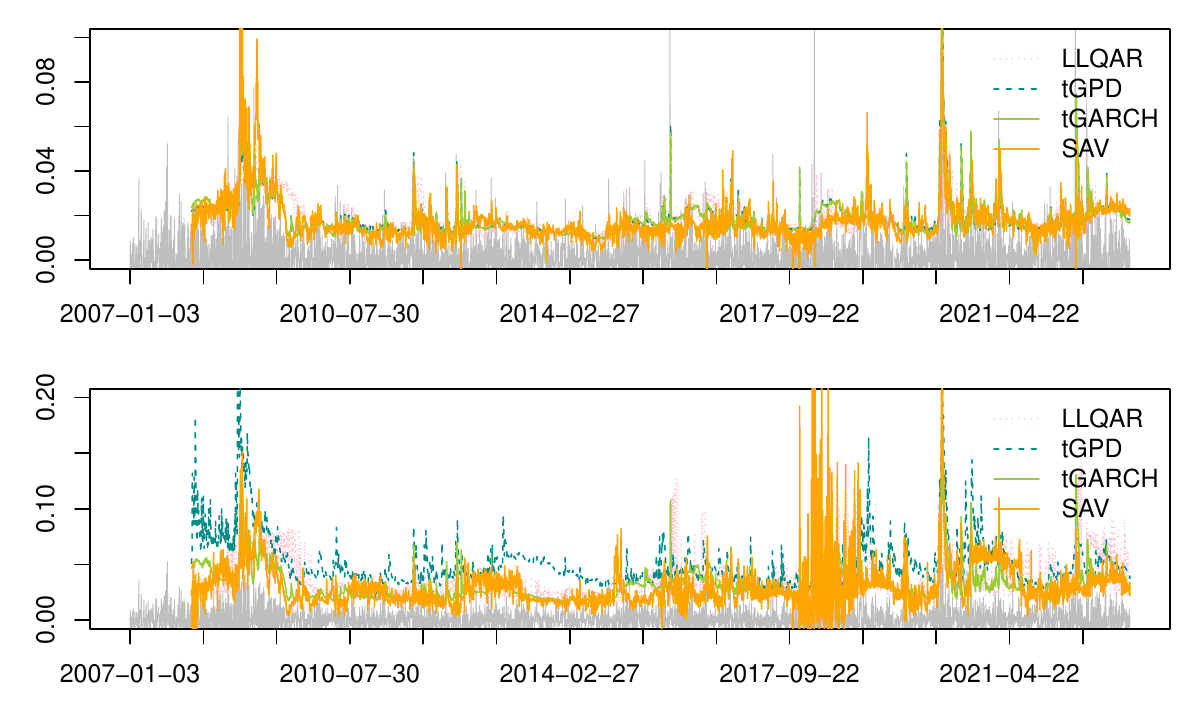}
       
      \caption{\textbf{Top}: Plot of $95\%$ (top) and $99\%$(bottom) VaR of different methods of WMT with AS method.}
    \end{figure}
    
    \begin{figure}[H]
    \centering
     
       \includegraphics[width=4.5in]{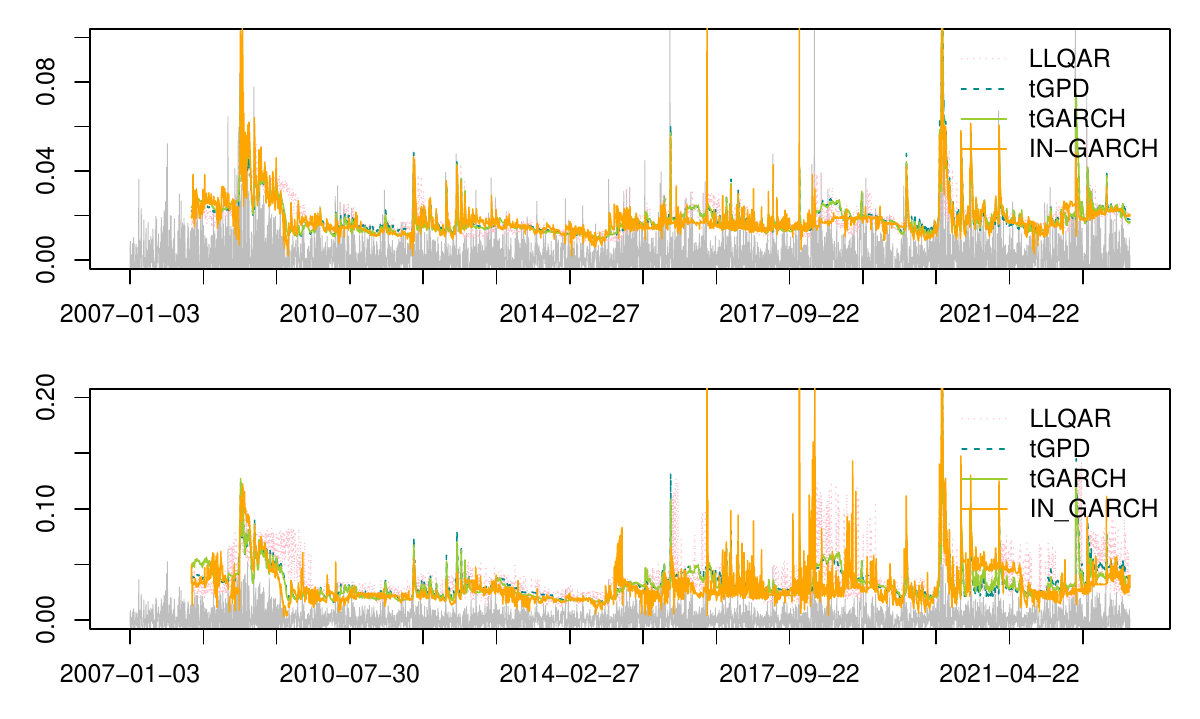}
       
      \caption{\textbf{Top}: Plot of $95\%$ (top) and $99\%$(bottom) VaR of different methods of WMT with In-direct-GARCH method.}
    \end{figure}

    \begin{figure}[H]
    \centering
     
       \includegraphics[width=4.5in]{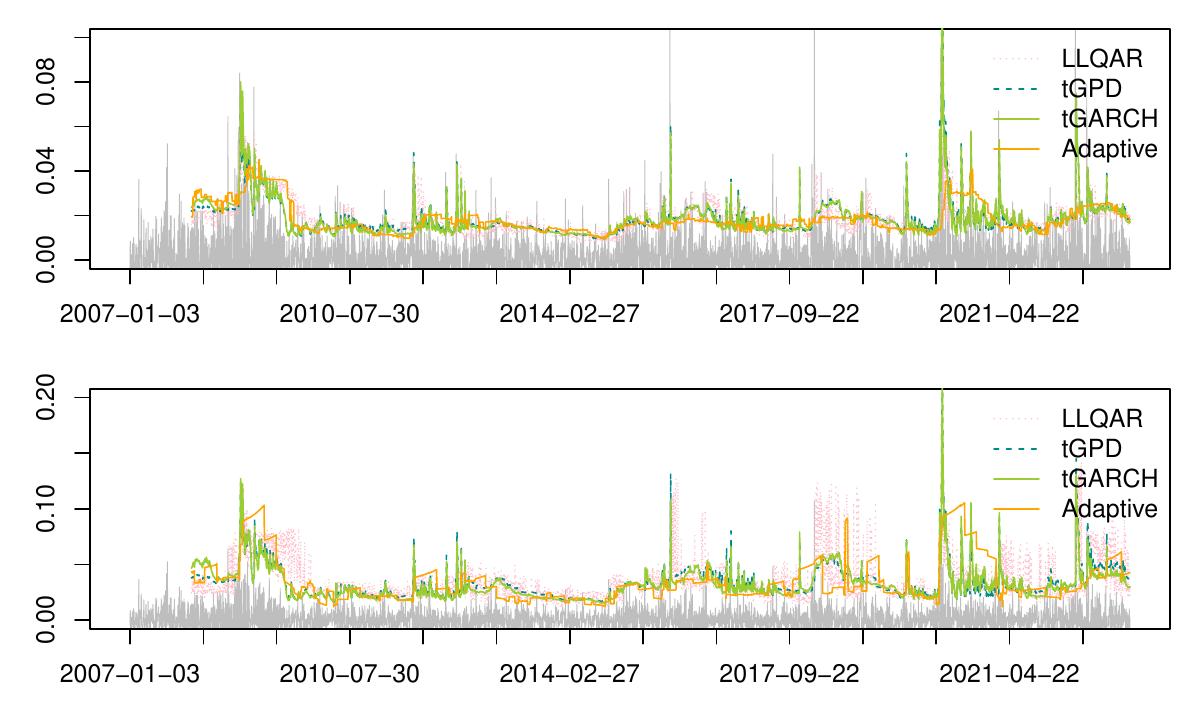}
       
      \caption{\textbf{Top}: Plot of $95\%$ (top) and $99\%$(bottom) VaR of different methods of WMT with Adaptive method.}
    \end{figure}

\end{document}